\documentclass[sigconf,authorversion,nonacm]{acmart}

\makeatletter
\def\@ACM@checkaffil{
    \if@ACM@instpresent\else
    \ClassWarningNoLine{\@classname}{No institution present for an affiliation}%
    \fi
    \if@ACM@citypresent\else
    \ClassWarningNoLine{\@classname}{No city present for an affiliation}%
    \fi
    \if@ACM@countrypresent\else
        \ClassWarningNoLine{\@classname}{No country present for an affiliation}%
    \fi
}
\makeatother

\copyrightyear{2023}
\acmYear{2023}
\setcopyright{acmlicensed}\acmConference[WWW '23]{Proceedings of the ACM Web Conference 2023}{May 1--5, 2023}{Austin, TX, USA}
\acmBooktitle{Proceedings of the ACM Web Conference 2023 (WWW '23), May 1--5, 2023, Austin, TX, USA}
\acmPrice{15.00}
\acmDOI{10.1145/3543507.3583217}
\acmISBN{978-1-4503-9416-1/23/04}

\expandafter\def\expandafter\UrlBreaks\expandafter{\UrlBreaks
    \do\a\do\b\do\c\do\d\do\e\do\f\do\g\do\h\do\i\do\j%
    \do\k\do\l\do\m\do\n\do\o\do\p\do\q\do\r\do\s\do\t%
    \do\u\do\v\do\w\do\x\do\y\do\z\do\A\do\B\do\C\do\D%
    \do\E\do\F\do\G\do\H\do\I\do\J\do\K\do\L\do\M\do\N%
    \do\O\do\P\do\Q\do\R\do\S\do\T\do\U\do\V\do\W\do\X%
    \do\Y\do\Z\do\/\do-}

\usepackage{breakurl}
\usepackage{subfigure}

\usepackage{float}
\usepackage{url}
\usepackage{xcolor}
\usepackage{balance}
\usepackage{xspace}
\usepackage{amsmath}
\usepackage{hyperref}
\usepackage{amsfonts}%
\usepackage{multirow}
\usepackage{graphicx}
\usepackage{booktabs}
\usepackage{caption}
\usepackage{array}
\usepackage{enumitem}
\setlist[itemize]{left=0pt}
\setlist[enumerate]{left=0pt}

\usepackage{color, colortbl}
\usepackage[first=0,last=9]{lcg}

\definecolor{Gray}{gray}{0.9}
\definecolor{LightCyan}{rgb}{0.88,1,1}

\newcommand{\etal}{\textit{et al.}\xspace}
\newcommand{\increaseR}{advantage\xspace}
\newcommand{\fixed}[1]{{\color{black}{#1}}}

\usepackage{lipsum}

\newcommand\blfootnote[1]{%
  \begingroup
  \renewcommand\thefootnote{}\footnote{#1}%
  \addtocounter{footnote}{-1}%
  \endgroup
}

\newcommand{\numdepositAttackers}{{$172$}\xspace}
\newcommand{\numwithdrawAttackers}{{$141$}\xspace}
\newcommand{\numtotalmixerAttackers}{{$205$}\xspace}
\newcommand{\totalDepositMixerAttackers}{{{$194{,}448$}}\xspace}
\newcommand{\totalWithdrawMixerAttackers}{{{$3{,}523.4$}}\xspace}
\newcommand{\numtotalmixerBEVextractors}{{$2{,}595$}\xspace}
\newcommand{\numdepositBEVextractors}{{$2{,}376$}\xspace}
\newcommand{\numwithdrawBEVextractors}{{$545$}\xspace}
\newcommand{\totalDepositMixerBEV}{{$115{,}980.5$\xspace}}
\newcommand{\totalWithdrawMixerBEV}{{$45{,}536.8$}\xspace}
\newcommand{\totalTCASReduced}{{$27.34\%$}\xspace}
\newcommand{\totalTNASReduced}{{$46.02\%$}\xspace}
\newcommand{\totalDepositofAttackersAndBEVextractors}{{$412.87$M}\xspace}

\newcommand{\crawlstartdate}{December 16th, 2019\xspace}
\newcommand{\crawlenddate}{{October 1st, 2022}\xspace}
\newcommand{\sanctiondate}{August 8th, 2022\xspace}
\newcommand{\crawlendBscdBlock}{{$21{,}800{,}000$}\xspace}
\newcommand{\tcSanctionStartBlock}{{$15{,}304{,}706$}\xspace}
\newcommand{\crawlendEthdBlock}{{$15{,}650{,}000$}\xspace}
\newcommand{\amstartdate}{December 18th, 2020\xspace}

\newcommand{\specialcell}[2][l]{%
  \begin{tabular}[#1]{@{}l@{}}#2\end{tabular}}

\newcommand{\observedAS}{\ensuremath{\xspace{\mathbf{OAS}_\mathbf{P}(t)}}\xspace}
\newcommand{\trueAS}{\ensuremath{\xspace{\mathbf{TAS}_\mathbf{P}(t)}}\xspace}
\newcommand{\simplifiedAS}{\ensuremath{\xspace{\mathbf{SAS}_\mathbf{P}(t)}}\xspace}

\newcommand{\sas}[1]{\ensuremath{\xspace{\mathbf{SAS}^{(#1)}_\mathbf{P}(t)}}\xspace}

\hypersetup{
    colorlinks=true,
    linkcolor=blue,
    filecolor=magenta,
    urlcolor=cyan,
    pdfpagemode=FullScreen,
}

\usepackage{acronym}

\acrodef{ZKP}[ZKP]{Zero-knowledge proof}
\newcommand{\ZKP}{\ac{ZKP}\xspace}

\acrodef{DeFi}[DeFi]{Decentralized Finance}
\newcommand{\DeFi}{\ac{DeFi}\xspace}

\acrodef{CeFi}[CeFi]{Centralized Finance}

\acrodef{PoS}[PoS]{Proof-of-Stake}

\acrodef{PoW}[PoW]{Proof-of-Work}

\acrodef{ENS}[ENS]{Ethereum Name Service}
\newcommand{\ENS}{\ac{ENS}\xspace}

\acrodef{TC}[TC]{Tornado.Cash}
\newcommand{\TC}{\ac{TC}\xspace}

\acrodef{TN}[TN]{Typhoon.Network}
\newcommand{\TN}{\ac{TN}\xspace}

\acrodef{TP}[TP]{Typhoon.Cash}
\newcommand{\TP}{\ac{TP}\xspace}

\acrodef{CEX}[CEX]{Centralized Exchange}
\newcommand{\CEX}{\ac{CEX}\xspace}
\newcommand{\CEXes}{\acp{CEX}\xspace}

\acrodef{DEX}[DEX]{Decentralized Exchange}

\newcommand{\DEXes}{\acp{DEX}\xspace}

\acrodef{P2P}[P2P]{peer-to-peer}
\newcommand{\PtP}{\ac{P2P}\xspace}

\acrodef{DApp}[DApp]{Decentralized Application}
\newcommand{\DApp}{\ac{DApp}\xspace}
\newcommand{\DApps}{\acp{DApp}\xspace}

\acrodef{MEV}[MEV]{Miner Extractable Value}
\newcommand{\MEV}{\ac{MEV}\xspace}

\acrodef{BEV}[BEV]{Blockchain Extractable Value}
\newcommand{\BEV}{\ac{BEV}\xspace}

\acrodef{FaaS}[FaaS]{Front-running as a Service}
\newcommand{\FaaS}{\ac{FaaS}\xspace}

\acrodef{OFAC}[OFAC]{Office of Foreign Assets Control}
\newcommand{\OFAC}{\ac{OFAC}\xspace}

\acrodef{AM}[AM]{anonymity mining}
\newcommand{\AM}{\ac{AM}\xspace}

\acrodef{BSC}[BSC]{Binance Smart Chain}
\newcommand{\BSChain}{\ac{BSC}\xspace}

\acrodef{ETH}[ETH]{Ethereum}
\newcommand{\ETHchain}{\ac{ETH}\xspace}

\acrodef{CLI}[CLI]{command line interface}
\newcommand{\CLI}{\ac{CLI}\xspace}

\begin{document}
\title{On How Zero-Knowledge Proof Blockchain Mixers Improve, and Worsen User Privacy}


\settopmatter{authorsperrow=4}

\author{Zhipeng Wang}
\affiliation{
  \institution{Imperial College London}
}

\author{Stefanos Chaliasos}
\affiliation{
  \institution{Imperial College London}
}

\author{Kaihua Qin}
\affiliation{
  \institution{Imperial College London}
  \institution{UC Berkeley RDI}
}

\author{Liyi Zhou}
\affiliation{
  \institution{Imperial College London}
  \institution{UC Berkeley RDI}
}

\author{Lifeng Gao}
\affiliation{
  \institution{Imperial College London}
}

\author{Pascal Berrang}
\affiliation{
  \institution{University of Birmingham}
}

\author{Benjamin Livshits}
\affiliation{
  \institution{Imperial College London}
}

\author{Arthur Gervais}
\affiliation{
  \institution{University College London}
  \institution{UC Berkeley RDI}
}



\renewcommand{\shortauthors}{Wang et al.}




\newcommand{\newcontent}[1]{{\color{red}#1}}
\newcommand{\pparagraph}[1]{\smallskip \indent \textbf{#1.}}

\newcommand{\alice}{\player{A}}
\newcommand{\bob}{\player{B}}
\newcommand{\carol}{\player{C}}
\newcommand{\dave}{\player{D}}
\newcommand{\zoe}{\player{Z}}

\newcommand{\hashs}{\ensuremath{\mathsf{H}_{\mathsf{s}}}\xspace}
\newcommand{\hashp}{\ensuremath{\mathsf{H}_{\mathsf{p}}}\xspace}
\newcommand{\hashf}{\ensuremath{\mathsf{H}}\xspace}
\newcommand{\mc}{}
\newcommand{\sparam}{\ensuremath{1^\lambda}\xspace}
\newcommand{\define}{\ensuremath{:=}\xspace}
\newcommand{\bit}{b\xspace}

\newcommand{\var}[1]{\texttt{#1}}
\newcommand{\fun}[1]{\textsc{#1}}
\newcommand{\player}[1]{\textsf{#1}}
\newcommand{\key}[1]{\mathsf{#1}}

\newcommand{\accGen}{\fun{ACC.Gen}\xspace} 
\newcommand{\accEval}{\fun{ACC.Eval}\xspace} 
\newcommand{\accWit}{\fun{ACC.Wit}\xspace} 

\newcommand{\desc}{\key{desc}\xspace}

\newcommand{\sokGen}{\fun{SoK.Gen}\xspace} 
\newcommand{\sokSign}{\fun{SoK.Sign}\xspace} 
\newcommand{\sokVerify}{\fun{SoK.Verify}\xspace} 

\newcommand{\crs}{\key{crs}\xspace}

\newcommand{\upkGen}{\fun{Gen}\xspace}
\newcommand{\upkUpdate}{\fun{Update}\xspace}
\newcommand{\upkVKP}{\fun{VerifyKP}\xspace}
\newcommand{\upkVU}{\fun{VerifyUpdate}\xspace}
\newcommand{\upkKeyExtract}{\fun{KeyExtraction}\xspace}

\newcommand{\serialgen}{\fun{SerialGen}\xspace}
\newcommand{\mint}{\fun{Mint}\xspace}
\newcommand{\accountgen}{\fun{AccountGen}\xspace}
\newcommand{\checkaccount}{\fun{CheckAccount}\xspace}
\newcommand{\spend}{\fun{Spend}\xspace}
\newcommand{\isspent}{\fun{IsSpent}\xspace}

\newcommand{\blockchainstate}{\mathbb{S}}
\newcommand{\serialnum}{\mathsf{s}\xspace}
\newcommand{\cn}{\mathsf{cn}\xspace}
\newcommand{\cnk}{\mathsf{cnk}\xspace}
\newcommand{\act}{\mathsf{act}\xspace}
\newcommand{\ask}{\mathsf{ask}\xspace}
\newcommand{\CKo}{\mathsf{CK}_{out}\xspace}
\newcommand{\amount}{\key{amt}\xspace}

\newcommand{\addr}{\mathsf{a}\xspace}

\newcommand{\sk}{\key{sk}\xspace}
\newcommand{\pk}{\key{pk}\xspace}


\newcommand{\pp}{\key{pp}\xspace}

\newcommand{\val}{\ensuremath{v}\xspace}

\newcommand{\sender}{payer\xspace}
\newcommand{\senders}{payers\xspace}
\newcommand{\receiver}{payee\xspace}
\newcommand{\receivers}{payees\xspace}

\newcommand{\Security}{Balance security\xspace}
\newcommand{\security}{balance security\xspace}
\newcommand{\deployability}{deployability\xspace}
\newcommand{\Deployability}{Deployability\xspace}
\newcommand{\privacy}{fungibility\xspace}
\newcommand{\Privacy}{Fungibility\xspace}

\newcommand{\sample}{\ensuremath{\xleftarrow{\$}}\xspace}
\newcommand{\coins}{\ensuremath{\xspace\texttt{coin}}\xspace}
\newcommand{\ETH}{\ensuremath{\xspace\texttt{ETH}}\xspace}
\newcommand{\DAI}{\ensuremath{\xspace\texttt{DAI}}\xspace}
\newcommand{\USDC}{\ensuremath{\xspace\texttt{USDC}}\xspace}
\newcommand{\BSC}{\ensuremath{\xspace\texttt{BSC}}\xspace}
\newcommand{\BNB}{\ensuremath{\xspace\texttt{BNB}}\xspace}
\newcommand{\MATIC}{\ensuremath{\xspace\texttt{MATIC}}\xspace}

\newcommand{\BLND}{\ensuremath{\xspace\texttt{BLND}}\xspace}

\newcommand{\BTC}{\ensuremath{\xspace\texttt{BTC}}\xspace}
\newcommand{\USD}{\ensuremath{\xspace\texttt{USD}}\xspace}

\newcommand{\TORN}{\ensuremath{\xspace\texttt{TORN}}\xspace}
\newcommand{\pTORN}{\ensuremath{\xspace\texttt{pTORN}}\xspace}
\newcommand{\PHOON}{\ensuremath{\xspace\texttt{PHOON}}\xspace}
\newcommand{\wBTC}{\ensuremath{\xspace\texttt{wBTC}}\xspace}
\newcommand{\cDAI}{\ensuremath{\xspace\texttt{cDAI}}\xspace}
\newcommand{\USDT}{\ensuremath{\xspace\texttt{USDT}}\xspace}
\newcommand{\MIC}{\ensuremath{\xspace\texttt{MIC}}\xspace}
\newcommand{\crDAI}{\ensuremath{\xspace\texttt{crDAI}}\xspace}
\newcommand{\xcoin}{\ensuremath{\gamma}\xspace}

\newcommand{\fromAddr}{\ensuremath{\xspace\texttt{from}}\xspace}
\newcommand{\toAddr}{\ensuremath{\xspace\texttt{to}}\xspace}
\newcommand{\Amount}{\ensuremath{\xspace\texttt{amt}}\xspace}
\newcommand{\blockNumber}{\ensuremath{\xspace\texttt{bn}}\xspace}
\newcommand{\Addr}{\ensuremath{\xspace\texttt{addr}}\xspace}

\newcommand{\swap}{\ensuremath{\xspace\texttt{Swap}}\xspace}
\newcommand{\publish}{\fun{Publish}\xspace}
\newcommand{\store}{\fun{Store}\xspace}


\newcommand{\openc}{\fun{OpenCh}\xspace}
\newcommand{\payc}{\fun{PayCh}\xspace}

\newcommand{\myeqdef}{\mathrel{\stackrel{\makebox[0pt]{\mbox{\normalfont\tiny def}}}{=}}}
\newcommand{\myeqtest}{\mathrel{\stackrel{\makebox[0pt]{\mbox{\normalfont\tiny $?$}}}{=}}}

\newcommand{\ZZl}{\ensuremath{\mathbb{Z}_q\xspace}}
\newcommand{\GG}{\ensuremath{\mathbb{G}\xspace}}
\newcommand{\PP}{\ensuremath{\mathbb{P}\xspace}}
\newcommand{\primeorder}{\ensuremath{q}\xspace}

\newcommand{\spendTx}{\fun{spendTx}\xspace}
\newcommand{\verifyTx}{\fun{VerifyTx}\xspace}
\newcommand{\amt}{\ensuremath{\gamma}\xspace}
\newcommand{\ledger}{\ensuremath{\mathcal{L}}\xspace}
\newcommand{\ledgertime}{\ensuremath{\ledger[\textsc{time}]}\xspace}

\newcommand{\com}{\mathsf{com}\xspace}
\newcommand{\range}{\ensuremath{\pi_{\mathsf{amt}}}\xspace}
\newcommand{\rangeprime}{\ensuremath{\Pi'_\textit{rge}}\xspace}
\newcommand{\rangeprimeprime}{\ensuremath{\Pi''_\textit{rge}}\xspace}
\newcommand{\ptime}{\ensuremath{\Pi\textit{-time}}\xspace}

\newcommand{\generator}{\ensuremath{g}\xspace}
\newcommand{\comgen}{\ensuremath{\mathcal{H}}\xspace}
\newcommand{\keyimage}{\ensuremath{\mathcal{I}}\xspace}
\newcommand{\dualkeyimage}{\ensuremath{\mathcal{J}}\xspace}
\newcommand{\viewkey}{\var{viewKey}\xspace}
\newcommand{\spendkey}{\var{spendKey}\xspace}
\newcommand{\otspendkey}{\var{otSpendKey}\xspace}
\newcommand{\saaddress}{\var{saAddr}\xspace}
\newcommand{\otaddress}{\var{otAddr}\xspace}
\newcommand{\hint}{\var{hint}\xspace}

\newcommand{\Weighted}{\mathsf{Weighted}\xspace}
\newcommand{\tr}{\mathsf{tr}\xspace}
\newcommand{\tx}{\mathsf{tx}\xspace}
\newcommand{\share}[1]{\ensuremath{[#1]}}
\newcommand{\set}[1]{\ensuremath{\{#1\}}\xspace}
\newcommand{\dtlc}{\textbf{DTLC}\xspace}
\newcommand{\ctlv}{\textbf{CLTV}\xspace}
\newcommand{\msg}{m\xspace}
\newcommand{\dig}{\xspace m \xspace}

\newcommand{\directtx}{\var{tx}\xspace}
\newcommand{\internaltx}{\var{itx}\xspace}
\newcommand{\mergetx}{\var{mtx}\xspace}
\newcommand{\checkedtx}{\var{ctx}\xspace}

\newcommand{\deptx}{\var{dtx}\xspace}
\newcommand{\reftx}{\var{rtx}\xspace}
\newcommand{\paytx}{\var{otx}\xspace}
\newcommand{\condtx}{\var{ctx}\xspace}

\newcommand{\lightcolor}{light blue\xspace}
\newcommand{\hl}[1]{\color{cyan}#1\color{black}}

\newcommand{\keygen}{\fun{KeyGen}\xspace}
\newcommand{\sign}{\fun{Sign}\xspace}
\newcommand{\verify}{\fun{Verify}\xspace}
\newcommand{\link}{\fun{Link}\xspace}
\newcommand{\tprove}{\fun{TProve}\xspace}
\newcommand{\tverify}{\fun{TVerify}\xspace}
\newcommand{\commit}{\fun{Commit}\xspace}
\newcommand{\open}{\fun{Open}\xspace}
\newcommand{\prove}{\fun{RProve}\xspace}
\newcommand{\rverify}{\fun{RVerify}\xspace}
\newcommand{\signoracle}{\ensuremath{\mathcal{SO}}\xspace}
\newcommand{\attacker}{\ensuremath{\mathcal{A}}\xspace}
\newcommand{\tcom}{\fun{Com}\xspace}
\newcommand{\trange}{\ensuremath{\Pi\textit{-tme}}\xspace}
\newcommand{\tproverel}{\fun{TProveRelation}\xspace}
\newcommand{\tverifyrel}{\fun{TVerifyRelation}\xspace}
\newcommand{\rcom}{\fun{Com}\xspace}
\newcommand{\rprove}{\fun{RProve}\xspace}

\newcommand{\gensaaddr}{\fun{GenSaAddr}\xspace}
\newcommand{\genotaddr}{\fun{GenOtAddr}\xspace}
\newcommand{\sacheckaddr}{\fun{CheckOtAddr}\xspace}
\newcommand{\sasetkey}{\fun{OtSpendKey}\xspace}

\newcommand{\twogensaaddr}{\fun{2of2GenSaAddr}\xspace}
\newcommand{\tworssign}{\fun{2of2RSSign}\xspace}
\newcommand{\verifyshare}{\fun{VerifyShare}}
\newcommand{\tworssigncond}{\fun{2of2RSSignCond}\xspace}
\newcommand{\tworssignone}{\fun{1of2RSSign}\xspace}
\newcommand{\condpay}{\fun{CondPay}\xspace}
\newcommand{\multicondpay}{\fun{MultiCondPay}\xspace}

\newcommand{\rskeygen}{\fun{RSKeyGen}\xspace}
\newcommand{\rssign}{\fun{RSSign}\xspace}
\newcommand{\rsverify}{\fun{RSVerify}\xspace}
\newcommand{\rssigma}{\ensuremath{\sigma}\xspace}

\newcommand{\createtime}{\fun{CreateTL}\xspace}

\newcommand{\zkprove}{\fun{ZKProve}}
\newcommand{\zkverify}{\fun{ZKVerify}}
\newcommand{\statement}{\textit{st}}

\newcommand{\TODOP}[1]{\ifdraft \textbf{TODO[P]: #1}  \fi}
\newcommand{\TODOR}[1]{\ifdraft \textbf{TODO[R]: #1}  \fi}

\newcommand{\rowseparation}{0.6}
\newcommand{\borderextra}{2}
\newcommand{\firstcolumn}{4cm}
\newcommand{\middleonecolumn}{2cm}
\newcommand{\secondcolumn}{8cm}
\newcommand{\middletwocolumn}{4cm}
\newcommand{\betweenonetwocolumn}{6cm}
\newcommand{\secondcolumnplus}{8cm}

\newcommand{\posa}{0 cm}
\newcommand{\posb}{4 cm}
\newcommand{\posc}{8 cm}
\newcommand{\posd}{12 cm}
\newcommand{\pose}{16 cm}
\newcommand{\posab}{2 cm}
\newcommand{\posbc}{6 cm}
\newcommand{\poscd}{10 cm}
\newcommand{\posde}{14 cm}

\newcommand{\firstrow}{-1*\rowseparation cm}
\newcommand{\secondrow}{-2*\rowseparation cm}
\newcommand{\thirdrow}{-3*\rowseparation cm}
\newcommand{\fourthrow}{-4*\rowseparation cm}
\newcommand{\fifthrow}{-5*\rowseparation cm}
\newcommand{\sixthrow}{-6*\rowseparation cm}
\newcommand{\seventhrow}{-7*\rowseparation cm}
\newcommand{\eighthrow}{-8*\rowseparation cm}
\newcommand{\ninethrow}{-9*\rowseparation cm}
\newcommand{\tenthrow}{-10*\rowseparation cm}
\newcommand{\eleventhrow}{-11*\rowseparation cm}
\newcommand{\twelvethrow}{-12*\rowseparation cm}
\newcommand{\thirteenthrow}{-13*\rowseparation cm}
\newcommand{\fourteenthrow}{-14*\rowseparation cm}
\newcommand{\fifteenthrow}{-15*\rowseparation cm}
\newcommand{\sixteenthrow}{-16*\rowseparation cm}
\newcommand{\seventeenthrow}{-17*\rowseparation cm}
\newcommand{\eighteenthrow}{-18*\rowseparation cm}
\newcommand{\nineteenthrow}{-19*\rowseparation cm}

\newcommand{\says}[2]{\noindent\textcolor{orange}{\textbf{#1 says: }}\textcolor{blue}{#2.} \xspace}
\newcommand{\setup}{$\fun{Setup(\secparam)}$ \xspace}
\newcommand{\poly}[1]{\ensuremath{\mathsf{poly}(#1)}}
\newcommand{\bset}{\ensuremath{\{0,1\}}}
\newcommand{\scalargroup}{\ensuremath{\mathbb{Z}_{\primeorder}}}

\newcommand{\adv}{\ensuremath{\mathcal{A}}\xspace}
\newcommand{\challenger}{\ensuremath{\mathcal{C}}\xspace}

\newcommand{\connect}{\fun{Connect}\xspace}
\newcommand{\replace}{\fun{Replace}\xspace}
\newcommand{\replaceS}{\fun{ReplaceS}\xspace}

\newcommand{\simplify}{\fun{Simp}\xspace}
\newcommand{\simplifyS}{\fun{SF}\xspace}

\newcommand{\merge}{\fun{Merge}\xspace}
\newcommand{\simplified}{\fun{Simp}\xspace}

\begin{abstract}\blfootnote{This paper is accepted at the ACM Web Conference 2023 (WWW '23).}
\ZKP \emph{mixers} are one of the most widely-used blockchain privacy solutions, operating on top of smart contract-enabled blockchains. We find that ZKP mixers are tightly intertwined with the growing number of \DeFi attacks and \BEV extractions. Through coin flow tracing, we discover that \numtotalmixerAttackers blockchain attackers and \numtotalmixerBEVextractors \BEV extractors leverage mixers as their source of funds, while depositing a total attack revenue of \totalDepositofAttackersAndBEVextractors USD. Moreover, the US OFAC sanctions against the largest ZKP mixer, Tornado.Cash, have reduced the mixer's daily deposits by more than {$80\%$}.

Further, ZKP mixers advertise their level of privacy through a so-called anonymity set size, which similarly to $k$-anonymity allows a user to hide among a set of $k$ other users. Through empirical measurements, we, however, find that these anonymity set claims are mostly inaccurate. For the most popular mixers on \ETHchain and \BSChain, we show how to reduce the anonymity set size on average by \totalTCASReduced and \totalTNASReduced respectively. Our empirical evidence is also the first to suggest a differing privacy-predilection of users on ETH and BSC.

State-of-the-art ZKP mixers are moreover interwoven with the DeFi ecosystem by offering \emph{\AM} incentives, i.e., users receive monetary rewards for mixing coins. However, contrary to the claims of related work, we find that AM does not necessarily improve the quality of a mixer's anonymity set. Our findings indicate that AM attracts privacy-ignorant users, who then do not contribute to improving the privacy of other mixer users.
\end{abstract}


\maketitle

\section{Introduction}\label{sec:introduction}
It is well-known that non-privacy-focused permissionless blockchains, such as Bitcoin and Ethereum, offer pseudonymity rather than anonymity~\cite{androulaki2013evaluating,gervais2014privacy, conti2018survey}. While privacy-preserving blockchains~\cite{miers2013zerocoin,sasson2014zerocash,hinteregger2018monerotraceability} aim to protect their users' privacy, retrofitting a blockchain with privacy has proven challenging and remains an active research area~\cite{tang2020privacy,meiklejohn2018mobius,  bonneau2014mixcoin,valenta2015blindcoin,heilman2017tumblebit,tairi2019a2l, ruffing2014coinshuffle,ruffing2017coinshufflepp, maxwell2013coinjoin}. The solution space can be broadly divided into \emph{(i)} privacy-by-design blockchains and \emph{(ii)} add-on privacy solutions, which are retrofitted, e.g., as a \DApp on top of non-privacy-preserving blockchains.

\ZKP mixers, inspired by Zerocash~\cite{sasson2014zerocash}, are one of the most widely-used blockchain privacy solutions, where a user deposits a \emph{fixed} denomination of coins into a pool and later withdraws these coins to an address. The goal of \ZKP mixers is to break the \emph{linkability} between a deposit and a new withdrawal address. The most active \ZKP mixer on \ETHchain, \TC, reports an anonymity set size of {$51{,}286$} for its largest pool (i.e., 1 \ETH pool) on \sanctiondate. This number is simply derived from the count of equal user deposits and suggests that, given a withdrawal transaction, the corresponding deposit can be hidden among the~$51$K deposits. Moreover, to attract users, \ZKP mixers offer anonymity mining (AM) incentives, where users can receive rewards for mixing coins.

\ZKP mixers have also attracted the attention of centralized regulators. On \sanctiondate, the US Treasury’s \OFAC placed sanctions~\cite{US-Treasury-Sanctions} on \TC due to alleged facilitation of money laundering. To our knowledge, this is the first time that centralized regulators sanctioned a decentralized and open-source application.


In this work, \emph{(i)} we investigate to what degree adversarial actors use \ZKP mixer, \emph{(ii)} how the \OFAC sanctions affect mixer usage, \emph{(iii)} we challenge the mixer's reported anonymity set sizes through heuristic intersections, and attempt to validate our heuristics through public side-channel data, and \emph{(iv)} we investigate the privacy implications of anonymity mining.



We summarize our contributions as follows:


\noindent \textbf{1. Analyzing Multi-Blockchain \ZKP Mixers Usage:}
We empirically investigate through coin flow tracing the deposit and withdrawal behavior on the two most popular \ZKP mixers, \TC (on \ETHchain) and \TN (on \BSChain). For mixer withdrawals, we discover that \numwithdrawAttackers malicious addresses and \numwithdrawBEVextractors \BEV extractors withdraw coins from a mixer as the adversarial source of funds. For mixer deposits, we find that \numdepositAttackers malicious addresses and \numdepositBEVextractors \BEV extractors deposit a total of \totalDepositofAttackersAndBEVextractors USD into \TC (cf.~Section~\ref{sec:overview}).

\noindent \textbf{2. \OFAC Sanctions Impact on Mixers:} We are the first to analyze how \OFAC sanctions affect \ZKP mixers. We find that, although $487$ user addresses have still deposited $62.59$M~USD into \TC after the sanctions, the total daily \TC deposits have decreased by $83\%$. Additionally, we discover that more than $85\%$ post-sanction \TC  withdrawn assets are transferred to intermediary addresses before being sent to \CEXes or \DeFi platforms, which indicates that users likely attempt to bypass the platforms' censorship (cf.~Section~\ref{sec:OFAC-sanctions}). 



\noindent  \textbf{3. Anonymity Mining's Impact on Privacy:} We are the first to study and empirically evaluate the impact of AM in \ZKP mixers. Contrary to the claims of related work~\cite{le2020amr}, we find that AM does not always increase mixers' anonymity set size quality, because AM appears to attract privacy-ignorant users with a primary interest in mining rewards. After pruning privacy-ignorant user addresses, we find that the \emph{\increaseR} (cf.~Eq.~\ref{eq:avg-increase}) that an adversary links a withdrawer to the correct depositor rises from $7.00\%$ (before AM launch) to $13.50\%$ (after AM launch) on average (cf. Section~\ref{sec:incentivisedpools}).

\noindent  \textbf{4. Measuring Mixer Anonymity Set Size:} 
We propose five on-chain data heuristics to derive a more accurate mixer anonymity set size, than naively enumerating equal user deposits. Combining heuristics proves powerful, as our evaluation shows that an adversary can reduce the anonymity set size on average by \totalTCASReduced and \totalTNASReduced of \TC (on ETH) and \TN (on BSC) respectively. We are hence the first to provide quantitative evidence indicating a user behavior difference w.r.t.\ privacy on two non-privacy-preserving blockchains. Our results also show that the biggest anonymity set continues to attract privacy-aware users, similar to how liquidity attracts liquidity in financial exchanges (cf. Section~\ref{sec:measuring-as}).

\section{Background}\label{sec:background}

\subsection{Blockchain and Smart Contracts}
    Permissionless blockchains act as a distributed ledger on top of a \PtP network. Smart contracts are quasi Turing-complete programs that typically execute within a virtual machine and allow users to construct various applications. For instance, \DeFi is a financial ecosystem that runs autonomously on smart-contracts-enabled blockchains. The total locked value in \DeFi has reached over~{$41$B} USD at the time of writing. Many \DApps are inspired by and mirror traditional centralized finance systems, such as asset exchanges, lending and borrowing platforms, and margin trading systems~\cite{daian2019flash,zhou2021high,qin2021empirical,qin2021attacking, wang2022speculative}. A transaction can be used to transfer blockchain tokens or to trigger the execution of smart contract functions. The sender of a transaction pays for the cost of the entire smart contract execution caused by that transaction. 
    

    Transactions are propagated over a public \PtP or a private relay network, prior to being validated by miners. Miners hence have the unilateral power to determine the transaction order in their mined blocks, creating an information asymmetry that yields a financial gain, i.e., \MEV~\cite{daian2019flash}. Generalizing \MEV, non-mining traders can also manipulate the transaction order and front-run their victims by paying higher transaction fees to extract \emph{blockchain extractable value} (\BEV)~\cite{qin2022quantifying}. Related work~\cite{qin2022quantifying} indicates that the dominant \BEV activities include sandwich attacks~\cite{zhou2021high}, liquidations~\cite{qin2021empirical}, arbitrages~\cite{zhou2021a2mm}, and replay attacks~\cite{qin2022quantifying}.

\subsection{Mixing Services for DeFi} \label{subsec:mixing-service}

Mixing services allow users to mix their coins with other users in an effort to break linkability of addresses. 
The literature features various proposals for mixing service designs, which can be centralized~\cite{bonneau2014mixcoin, valenta2015blindcoin, heilman2017tumblebit, tairi2019a2l} or governed by smart contracts.

As \DeFi adoption increases and all transactions, balances, senders, and recipients are public, the demand for privacy in DeFi has led to the launch of ZKP mixers.
To date, the largest \ZKP mixer on Ethereum is \TC~\cite{Tornadocash}, which launched in December 2019. \TC operates four \ETH pools (i.e., \href{https://etherscan.io/address/0x12d66f87a04a9e220743712ce6d9bb1b5616b8fc}{0.1}, \href{https://etherscan.io/address/0x47CE0C6eD5B0Ce3d3A51fdb1C52DC66a7c3c2936}{1}, \href{https://etherscan.io/address/0x910cbd523d972eb0a6f4cae4618ad62622b39dbf}{10} and \href{https://etherscan.io/address/0xa160cdab225685da1d56aa342ad8841c3b53f291}{100}~\ETH pools) which support the deposit and withdrawal of a fixed amount of \ETH. When a user deposits a fixed amount of \ETH into a \TC pool, the user should safely back up a deposit \emph{note}; to withdraw, the user should provide the deposit \emph{note}, which needs to be verified by the \TC smart contract. \TC also supports the mixing of other tokens (e.g., \USDC, \USDT, etc), but most users appear to be mixing \ETH. The total \ETH deposited in \TC reached over~{$3.54$M}~\ETH\footnote{We adopt the coin prices on \href{https://coinmarketcap.com/}{CoinMarketCap} on \crawlenddate, e.g., $1$~\ETH = $1{,}330$~USD, $1$~\BNB = $285$~USD.}~($4.70$B~USD) at the time of writing.

AMR~\cite{le2020amr} is a new mixer design similar to \TC, but additionally rewards its users for their participation in the system. Such incentivization of paying rewards is similar to the currently popular liquidity mining, also called ``DeFi farming'', an attempt to attract more users. More users should translate to a larger anonymity set size, as AMR proclaims. Soon after AMR, \TC was updated to support anonymity mining~\cite{TornadoCashGovernanceProposal} to incentivize users to keep their deposited \ETH in mixer pools for a longer time period. \ZKP mixers can also run on other smart contract-enabled blockchains, e.g., \TP on ETH, \TN and \href{https://cyclone.xyz/bsc}{Cyclone} on \BSChain. 



\subsection{\OFAC Sanctions against \TC}
On \sanctiondate, the US Treasury’s \OFAC placed sanctions~\cite{US-Treasury-Sanctions, Cyber-related-Sanctions} on  \TC, due to alleged assistance of money laundering. \OFAC added the \TC website and related addresses to the ``Specially Designated Nationals And Blocked Persons'' (SDN) list. According to the sanctions, US citizens are no longer legally allowed to use the \TC website or involve any property or interest transactions with the addresses in the SDN list. To our knowledge, this is the first time that centralized regulators sanction decentralized applications. The sanctions caused a series of consequences. For instance, many \DeFi platforms (e.g., Uniswap), \FaaS platforms (e.g., Flashbots), and miners (e.g., Ethermine) choose to censor \TC-related transactions or addresses interacting with \TC~\cite{chainalysis-tc-sanctions-challenges}.

\section{System Model and Privacy Metrics}\label{sec:basic-model}
In this section, we outline our system and privacy metrics.

\subsection{System Model}\label{sec:system-model}
\noindent\textbf{Address:} Users have at least one public/private key-pair (corresponding to their \emph{address}), which controls cryptocurrency assets on a permissionless blockchain. To transfer or trade an asset, a user signs a transaction with its private key. Each transaction corresponds to an event with various publicly readable features, such as the time of day and the transaction fees. 


\begin{figure}[t]
    \centering
    \includegraphics[width=0.9\columnwidth]{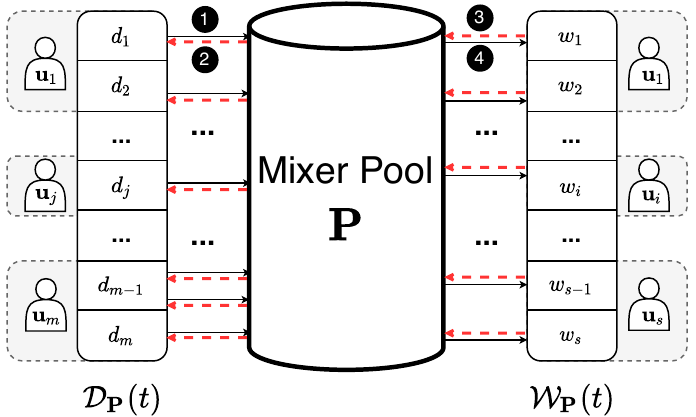}
    \caption{System Model for a mixer pool, where  $\mathcal{D}_\mathbf{P}(t)= \{d_1, ..., d_m\}$ and  $\mathcal{W}_\mathbf{P}(t)= \{w_1, ..., w_s\}$. `$\longrightarrow$' represents a transfer of \coins, and `{\color{red}$\dashleftarrow$}' represents a \emph{note} transfer.  {When a user $\mathbf{u}$ deposits \coins into pool $\mathbf{P}$ (in step~{1}), $\mathbf{u}$ receives a \emph{note} from $\mathbf{P}$ (in step~{2}). To withdraw, $\mathbf{u}$ 
   needs to provide \emph{note} to $\mathbf{P}$ (in step~{3}), and will receive \coins after $\mathbf{P}$ verifies \emph{note} (in step~{4}). A user can control multiple addresses. An address can be used to deposit or withdraw multiple times.}}

   \Description{System Model for a mixer pool, where  $\mathcal{D}_\mathbf{P}(t)= \{d_1, ..., d_m\}$ and  $\mathcal{W}_\mathbf{P}(t)= \{w_1, ..., w_s\}$. `$\longrightarrow$' represents a transfer of \coins, and `{\color{red}$\dashleftarrow$}' represents a \emph{note} transfer.  {When a user $\mathbf{u}$ deposits \coins into pool $\mathbf{P}$ (in step~{1}), $\mathbf{u}$ receives a \emph{note} from $\mathbf{P}$ (in step~{2}). To withdraw, $\mathbf{u}$ 
   needs to provide \emph{note} to $\mathbf{P}$ (in step~{3}), and will receive \coins after $\mathbf{P}$ verifies \emph{note} (in step~{4}). A user can control multiple addresses. An address can be used to deposit or withdraw multiple times.}}
    \label{fig:model1}
\end{figure}

\noindent\textbf{Coin Transfer:} A transfer of a \coins is a tuple  $\tr = (\blockNumber, \fromAddr, \toAddr, \Amount,$ $\coins)$, where $\blockNumber$ is the block number (i.e., timestamp), $\Amount$ is the amount of \coins that is transferred from the address $\fromAddr$ to $\toAddr$. 

\noindent\textbf{Coin Flow:} A chain of transfers of \coins between addresses.



\noindent\textbf{Link:} Two addresses $\addr_1$ and $\addr_2$ belong to the same user are linked. Denoted as $\link(\addr_1, \addr_2) = 1$.


\noindent\textbf{Cluster:} A cluster is a set of mutually-linked addresses.


\noindent\textbf{Mixer Pool:} A mixer pool, denoted as $\mathbf{P}$, is an aggregation of cryptocurrency assets governed by smart contracts (cf.\ Fig.~\ref{fig:model1}). Users can only deposit and withdraw a specific cryptocurrency \coins. To avoid that deposit/withdrawal asset amounts leak privacy, mixer pools typically only accept a fixed currency denomination. The proper use of a mixer pool $\mathbf{P}$ requires choosing one address to deposit and another ideally unlinkable address to withdraw. 

A \emph{depositor} is an address to deposit \coins into $\mathbf{P}$, and a \emph{withdrawer} is an address to receive \coins from $\mathbf{P}$. At time $t$, given a pool $\mathbf{P}$, denote its depositor set as $\mathcal{D}_\mathbf{P}(t)$ and withdrawer set as $\mathcal{W}_\mathbf{P}(t)$.

\begin{figure}[t]
    \centering
    \includegraphics[width = \linewidth]{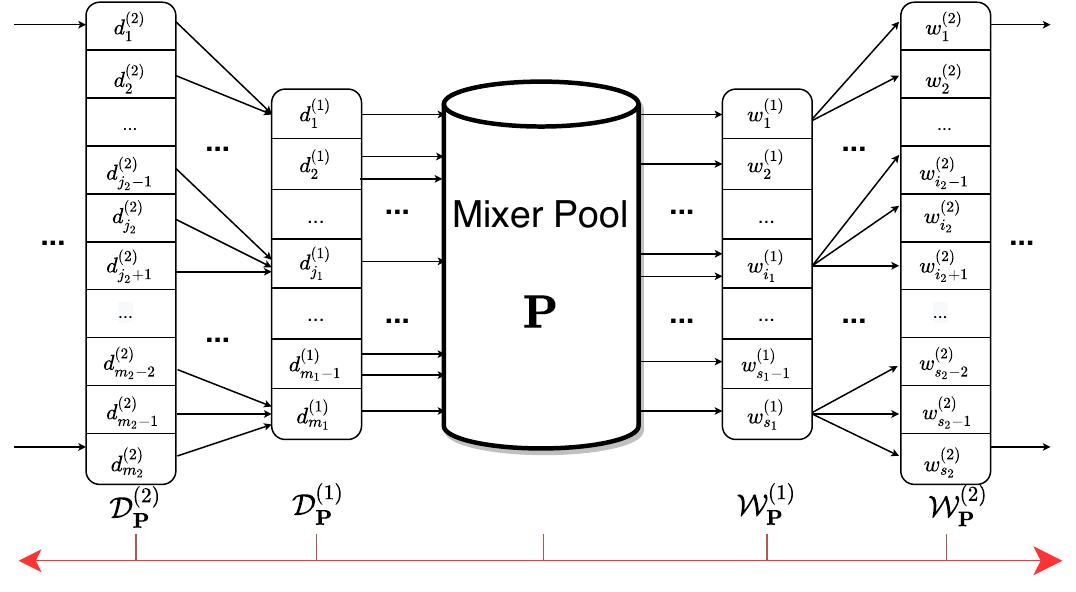}
    \caption{Extended model of a mixer pool. The mixer pool model in Fig.~\ref{fig:model1} can be extended to a model, which can cover depositors and withdrawers in longer distances.}
    \Description{Extended model of a mixer pool. Based on the extension of depositors and withdrawers, the mixer pool model in Fig.~\ref{fig:model1} can be extended to a model, which can cover depositors and withdrawers in longer distances.}
  \label{fig:model2}
\end{figure}

To track users' coin flows before and after interacting with a mixer pool, we extend the depositor and withdrawer set (cf.~Fig.~\ref{fig:model2}).

\smallskip\noindent\textbf{Depositors Extension:} At time $t$, we let $\mathcal{D}_\mathbf{P}(t) = \mathcal{D}^{(1)}_\mathbf{P}(t)$, and define the depositors in distance $n$ (where $n > 1$), $\mathcal{D}^{(n)}_\mathbf{P}(t)$, as the set of addresses that transfer \coins to the addresses in $\mathcal{D}^{(n-1)}_\mathbf{P}(t)$.

\smallskip\noindent\textbf{Withdrawers Extension:} At time $t$, we let $\mathcal{W}_\mathbf{P}(t) = \mathcal{W}^{(1)}_\mathbf{P}(t)$ and define the withdrawers in distance $n$ (where $n > 1$), $\mathcal{W}^{(n)}_\mathbf{P}(t)$, as the set of addresses that receive \coins from the addresses in $\mathcal{W}^{(n-1)}_\mathbf{P}(t)$.

\smallskip\noindent\textbf{Extended Mixer Pool:} Based on the extension of depositors and withdrawers, the mixer pool model in Fig.~\ref{fig:model1} can be extended to a model in Fig.~\ref{fig:model2}, which can cover depositors and withdrawers in {longer} distances.

We propose the following definitions to further describe mixers and summarize the key definitions in Table~\ref{tab:model-definitions}.

\smallskip\noindent\textbf{Address Balance in A Pool:} An address's balance is the amount of coins that an address holds \emph{in a pool} at a time $t$ (cf.~Eq.~\ref{eq:balance}).

\smallskip\noindent\textbf{Pool State:} A pool's state is the set of tuples constituted by all depositors, withdrawers, and their balances in $\mathbf{P}$, at time $t$ (cf.~Eq.~\ref{eq:poolstate}).

A pool $\mathbf{P}$'s state is determined by users' balances. For instance, if $d_1$ deposits once, $d_2$ deposits twice, and $w_1$ withdraws once in a $100$ \coins pool $\mathbf{P}_{100}$ before time $t$, then $\mathbf{P}_{100}$'s pool state is  $\mathbb{S}_{\mathbf{P}_{100}}(t) = \{(d_1, 100), (d_2, 200), (w_1, -100)\}$. If there exists a link between a depositor and a withdrawer in a pool $\mathbf{P}$, we can simplify the pools' state (cf.~Eq.~\ref{eq:simplified_pool}). For instance, if $\link(d_1, w_1) = 1$, then we can simplify the state as $\simplify \left(\mathbb{S}_{\mathbf{P}_{100}}(t),(d_1, w_1)\right) = \{(d_2, 200)\}$.

\begin{table*}[t]
\caption{System Model Definitions}
\footnotesize
\resizebox{\textwidth}{!}{
\begin{tabular}{llr|llr}
\toprule
\textbf{Name} 
& \textbf{Definition}
& \textbf{Eq.}

&\textbf{Name} 
& \textbf{Definition}
& \textbf{Eq.}

\\ 
\midrule

\rowcolor{Gray}
Coin Transfer 
& $\tr = (\blockNumber, \fromAddr, \toAddr, \amount, \coins) \ \text{where} \ \fromAddr \xrightarrow{\coins} \toAddr$
& \refstepcounter{equation}(\theequation)\label{eq:coin-transfer}

&\multirow{2}{*}{Coin Flow}
 & \multirow{2}{*}{\specialcell[l]{$\mathcal{F} = \left(\tr_1,..., \tr_n\right) \ \text{where}$ \\ $\tr_{i-1}.{\toAddr} = \tr_{i}.{\fromAddr} \ \text{and} \ \tr_{i-1}.{\blockNumber} \leq \tr_{i}.{\blockNumber}$}}
 & \multirow{2}{*}{\refstepcounter{equation}(\theequation)\label{eq:coin-flow}}\\[0.4cm]

Link
& $\link(\addr_1, \addr_2) = 1 \mbox{  $\Leftrightarrow\addr_1$ is linked to $\addr_2$}$ 
& \refstepcounter{equation}(\theequation)\label{eq:link}

&\multirow{2}{*}{Cluster}
& \multirow{2}{*}{\specialcell[l]{$\mbox{$\mathcal{C} = \{\addr_1, .., \addr_n\}$, $\forall \addr_i \in \mathcal{C}$, $\exists \addr_j \in \mathcal{C}\setminus\{\addr_i\},$}$ \\ $\mbox{satisfies $\link(\addr_i, \addr_j) = 1$}$}}
& \multirow{2}{*}{\refstepcounter{equation}(\theequation)\label{eq:cluster}}\\[0.4cm]

\rowcolor{Gray}

Pool Depositors  
& $\mathcal{D}_\mathbf{P}(t) = \{ \mathbf{d} \mid\mbox{$\mathbf{d}$ deposits \coins into $\mathbf{P}$ before $t$}\}$
& \refstepcounter{equation}(\theequation)\label{eq:pool-depositors}

&Pool Withdrawers  
& $\mathcal{W}_\mathbf{P}(t) = \{\mathbf{w} \mid\mbox{$\mathbf{w}$ withdraws \coins from $\mathbf{P}$ before $t$} \}$
& \refstepcounter{equation}(\theequation)\label{eq:pool-withdrawers}\\[0.3cm]

\multirow{2}{*}{Address Balance}
& \multirow{2}{*}{\specialcell[l]{$\mathbf{bal}_\addr(t) = \mathbf{u}_\addr(t) \times p - \mathbf{v}_\addr(t) \times p$, where $\mathbf{u}_\addr(t)$ and $\mathbf{v}_\addr(t)$ are \\
the numbers of $\addr$'s deposit and withdrawal, respectively.}} 
& \multirow{2}{*}{\refstepcounter{equation}(\theequation)\label{eq:balance}}

&Pool State
& $\mathbb{S}_\mathbf{P}(t) = \left\{(\addr, \mathbf{bal}_\addr(t)) \mid\addr \in \mathcal{D}_\mathbf{P}(t) \cup \mathcal{W}_\mathbf{P}(t) \right\}$
& \refstepcounter{equation}(\theequation)\label{eq:poolstate}\\[0.4cm]

\rowcolor{Gray}
\multirow{3}{*}{Merge}
& \multirow{3}{*}{\specialcell[l]{$\merge \left(\mathbb{S}_\mathbf{P}(t),(\addr_1, \addr_2)\right) = \{(\addr, \mathbf{bal}_\addr(t)) |\addr \in \mathcal{D}_\mathbf{P}(t) \cup \mathcal{W}_\mathbf{P}(t)$ \\ $\land \addr \not= \addr_1 \land \addr \not= \addr_2\} \cup \{(\addr_1, \mathbf{bal}_{\addr_1}(t) + \mathbf{bal}_{\addr_2}(t))\}$}} 

& \multirow{3}{*}{\refstepcounter{equation}(\theequation)\label{eq:merge}}

&\multirow{3}{*}{\specialcell[l]{Simplified Pool \\ State}}
& \multirow{3}{*}{\specialcell[l]{{If $\mathcal{S} = \emptyset$, $\simplified(\mathbb{S}_\mathbf{P}(t),\mathcal{S}) = \mathbb{S}_\mathbf{P}(t)$;}\\
Else: $\simplified(\mathbb{S}_\mathbf{P}(t),\mathcal{S})= \simplified \left(\merge(\mathbb{S}_\mathbf{P}(t), (\addr_i, \addr_{i+1})),\mathcal{S}^{'}\right)$ \\ $\mbox{where $S$ is a set of linked addresses and } \mathcal{S}^{'} = \mathcal{S}\setminus\{(\addr_i, \addr_{i+1})\}$}}
& \multirow{3}{*}{\refstepcounter{equation}(\theequation)\label{eq:simplified_pool}}\\[0.8cm]

\multirow{2}{*}{\specialcell[l]{Depositors \\ Extension}}
& \multirow{2}{*}{$\mathcal{D}^{(n)}_\mathbf{P}(t) = \{\addr \mid \exists \addr_{1}\in \mathcal{D}^{(n-1)}_\mathbf{P}(t), \addr \xrightarrow{\coins}  \addr_{1}\ \mbox{before}\ t\}$}
& \multirow{2}{*}{\refstepcounter{equation}(\theequation)\label{eq:depositors_extension}}

&\multirow{2}{*}{\specialcell[l]{Withdrawers \\ Extension}}
& \multirow{2}{*}{$\mathcal{W}^{(n)}_\mathbf{P}(t) = \{\addr \mid \exists \addr_{1}\in \mathcal{W}^{(n-1)}_\mathbf{P}(t), \addr_{1} \xrightarrow{\coins}  \addr\ \mbox{before}\ t\}$}
& \multirow{2}{*}{\refstepcounter{equation}(\theequation)\label{eq:withdrawers_extension}} \\[0.4cm]




\bottomrule
\end{tabular}
}
\label{tab:model-definitions}
\end{table*}

\subsection{Privacy Metrics}\label{sec:privacy-metrics}

Knowing the depositor set $\mathcal{D}_{\mathbf{P}}(t)$ of a pool $\mathbf{P}$ at time $t$, we define the \emph{observed} anonymity set and the \emph{true} anonymity set of the pool.

\smallskip\noindent\textbf{Observed Anonymity Set:} Given a mixer pool $\mathbf{P}$ at time $t$, the observed anonymity set \observedAS of a pool $\mathbf{P}$ is the set of unique deposit addresses, i.e., $\mathcal{D}_{\mathbf{P}}(t)$. 

\smallskip\noindent\textbf{True Anonymity Set:} At time $t$, the true anonymity set \trueAS of a pool $\mathbf{P}$ is the set of addresses with a positive deposit balance in the pool, i.e., the set of depositors whose deposited assets have not yet been completely withdrawn from the pool $\mathbf{P}$. 

Note that the true anonymity set might not be apparent from observing the blockchain data, because it is the mixer's intention to obfuscate the addresses depositing into the mixer pool. {However, an adversary can leverage on-chain data to compute a more ``realistic'' anonymity set, which can be more representative than \observedAS.}

\smallskip\noindent\textbf{Simplified Anonymity Set:} Given a mixer pool $\mathbf{P}$ at time $t$, the simplified anonymity set $\simplifiedAS$ is the set of depositors with a positive balance, which is computed by leveraging on-chain data to simplify the pool state. Note that $\simplifiedAS \subseteq \observedAS$.






\smallskip\noindent\textbf{Privacy Metric:}
The probability that an adversary without prior knowledge links a withdrawer (who withdraws at time $t$) to the correct depositor is $\mathsf{Adv}^{o}_{\adv}(t) = 1/\left|\mathbf{OAS}_\mathbf{P}(t)\right|$.
    
If the adversary can link a withdrawer $w$, to a target set of depositors $\mathbf{SAS}_\mathbf{P}(t)$, then the probability that the adversary links $w$ to the correct depositor is $\mathsf{Adv}^{s}_{\adv}(t) = 1/\left|\mathbf{SAS}_\mathbf{P}(t)\right|$.

 We further define $\mathsf{R}_\mathsf{Adv}$ as the increase of $\mathsf{Adv}^{s}_{\adv}(t)$ over $\mathsf{Adv}^{o}_{\adv}(t)$, to represent the \emph{\increaseR} that an adversary links a withdrawer to the correct depositor after simplifying the anonymity set (cf.~Eq.~\ref{eq:avg-increase}). 
\begin{equation}\label{eq:avg-increase}
\begin{aligned}
\mathsf{R}_\mathsf{Adv} = \frac{\mathsf{Adv}^{s}_{\adv}(t)-\mathsf{Adv}^{o}_{\adv}(t)}{\mathsf{Adv}^{o}_{\adv}(t)}
\end{aligned}
\end{equation}

\section{Empirical Mixer Activity}\label{sec:overview}
\begin{figure}[t]
    \centering
   \includegraphics[width=\columnwidth]{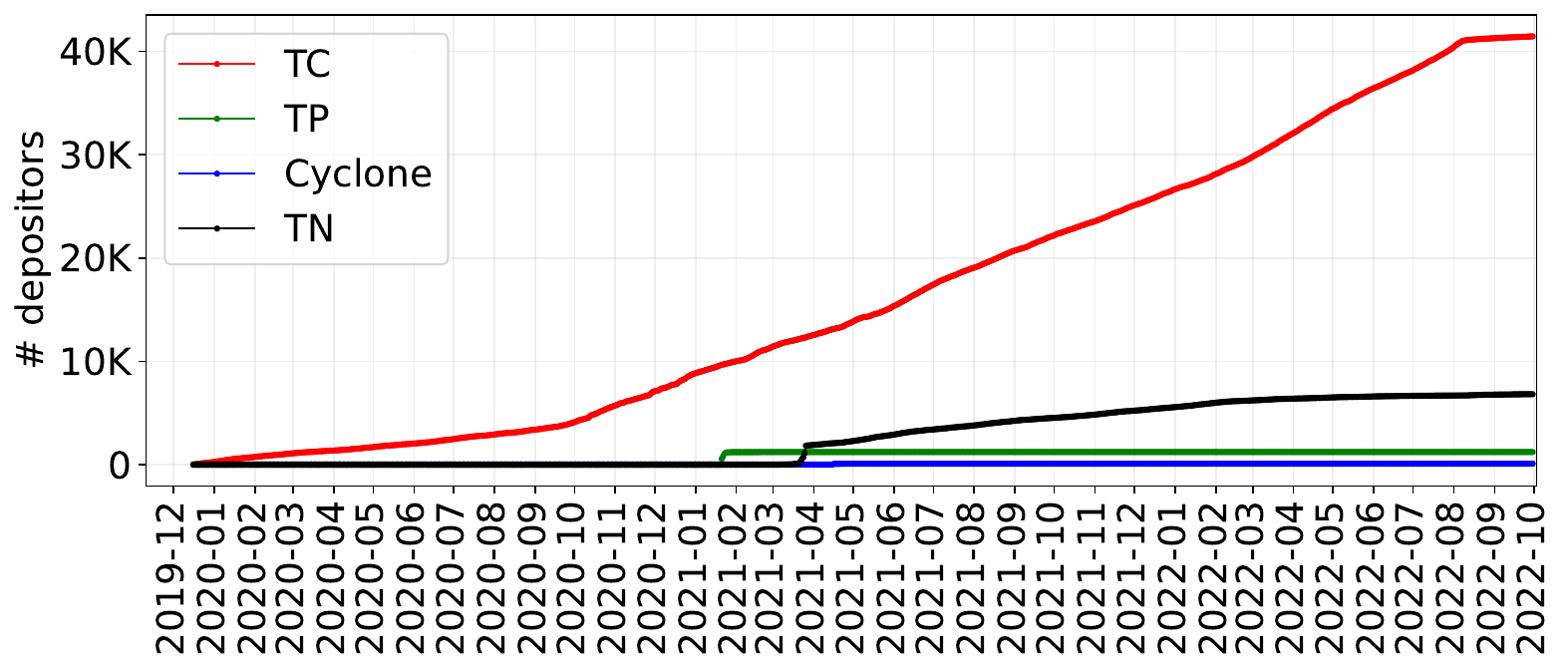}
\caption{\ZKP mixer depositors over time. {$41{,}441\thinspace (83.61\%)$} and {$6{,}814\thinspace (13.75\%)$} depositors appear in \TC and \TN, respectively.}
\Description{\ZKP mixer depositors over time. {$41{,}441\thinspace (83.61\%)$} and {$6{,}814\thinspace (13.75\%)$} depositors appear in \TC and \TN, respectively.}
\label{fig:mixer_depositor_over_time}
\end{figure}

To gather empirical insights into the activities of existing \ZKP mixers, we crawl the deposit, withdrawal events and transactions of the~$73$ pools on four \ZKP mixers: \TC, \TP, \TN and Cyclone, from \crawlstartdate (i.e., the inception time of \TC) to \crawlenddate. 
We observe that~{$97.36\%$} of the mixer users deposit assets into \TC and \TN, and that the number of \TP depositors has not changed since February,~2021 (cf.\ Fig.~\ref{fig:mixer_depositor_over_time}). Therefore, we focus on analyzing the two most active mixers, \TC and \TN.

We analyze the top four active pools in \TC ($0.1$, $1$, $10$ and~$100$ \ETH pools) and \TN ($0.1$, $1$, $10$ and $50$ \BNB pools). For \TC, we crawl the deposit and withdrawal events data from the Ethereum block $9{,}116{,}966$ (\crawlstartdate) to \crawlendEthdBlock (\crawlenddate). The \TC $1$ \ETH pool is the most active ({$51{,}770$} deposits and {$49{,}086$} withdrawals), while the TC $100$ \ETH pool has the smallest depositor and withdrawer set ({$6{,}433$} deposit and {$11{,}069$} withdraw addresses). The \TC pools accumulate deposits of {$3.54$M~\ETH ($4.70$B~USD)}. Moreover, from \TN's inception at \BSChain block $5{,}230{,}899$ (February 27th, 2021) until block \crawlendBscdBlock (\crawlenddate), we find that {$6{,}814$} addresses generate {$29{,}129$} deposits in the four \BNB pools, accumulating {$93{,}409.5$~\BNB ($26.62$M~USD)}.

\begin{table}[t]
\centering
\caption{Deposits/withdrawals in \TC \ETH and \TN \BNB pools.
}

\renewcommand\arraystretch{1.}
\resizebox{\columnwidth}{!}{
\begin{tabular}{lrrrrr}
\toprule
\bf Pool
& \multicolumn{1}{c}{\# Deposits}
& \multicolumn{1}{c}{\# Withdrawals}
& \multicolumn{1}{c}{\# Depositors}
& \multicolumn{1}{c}{\# Withdrawers}
\\
\midrule


TC 0.1 \ETH
&$26{,}069$ &$22{,}281$ &$11{,}941$ &$13{,}227$ \\ 

TC 1 \ETH
&$51{,}770$ &$49{,}086$ &$17{,}843$ &$23{,}592$ \\ 

TC 10 \ETH
&$45{,}238$ &$44{,}228$ &$16{,}227$ &$21{,}872$ \\ 

TC 100 \ETH
&$30{,}301$ &$29{,}553$ &$6{,}433$ &$11{,}069$ \\

\hline


TN 0.1 \BNB
&$10{,}485$ &$9{,}877$ &$3{,}972$ &$4{,}541$ \\ 

TN 1 \BNB
&$13{,}151$ &$12{,}901$ &$3{,}890$ &$4{,}362$ \\  

TN 10 \BNB
 &$4{,}886$ &$4{,}860$ &$1{,}675$ &$1{,}983$ \\ 

TN 50 \BNB
&$607$ &$604$ &$231$ &$288$ \\






\bottomrule
\end{tabular}
}
\label{tab: crawled_data}
\end{table}

\subsection{Depositors and Withdrawers}
The four \TC \ETH pools contain {$39{,}821$} depositors and {$61{,}026$} withdrawers, depositing {$88.82$~\ETH ($118$K~USD)} and withdrawing {$56.52$~\ETH ($75$K~USD)} on average. In each pool, the number of withdrawers is greater than depositors, indicating that a user may adopt multiple addresses to withdraw than to deposit. Moreover, {$58{,}998~(84.95\%)$} withdrawers have zero \ETH before receiving \ETH from \TC.

\smallskip\noindent\textbf{Cross-pool Mixer Usage.} Because a mixer pool only supports a fixed currency denomination, users may utilize multiple pools to mix arbitrary amounts of assets. We find that {$327$} depositors utilize all four \TC pools, and {$9{,}962$ ($25.02\%$)} deposit in more than one pool. Additionally, $60$ users withdraw from all four pools, and {$7{,}479~(12.26\%)$} use more than one pool to withdraw. Likewise, for \TN, we observe a slight increase in overlaps on both depositors ({$33$\%}) and withdrawers {($25$\%)} appearing in at least two pools. The overlap of pools may help an adversary to link addresses (cf.\  Section~\ref{sec:heuristics_all}).

\subsection{ZKP Mixer Coin Flow}
In addition to immediate depositors and withdrawers, we are also interested in the coins' wider flow to get their origins and destinations. For example, users move their coins from exchanges or DeFi platforms via intermediary addresses into and outside the mixer. 


To track where the deposited \ETH in \TC are transferred from and where the withdrawn \ETH are transferred to, we extend our pool model to cover depositors and withdrawers in distance $2$. We crawl the transaction history of user addresses before \crawlenddate. 


For each depositor $d^{(1)}$ in a \TC $p$ \ETH pool, we extract \emph{the most recent} transfers of $p$ \ETH that $d^{(1)}$ receives before depositing into \TC, and obtain the depositors in distance 2 that transfer \ETH to $d^{(1)}$. Similarly, we obtain the withdrawers in distance 2 by extracting \emph{the most recent} transfers of $p$ \ETH that the withdrawers in distance 1 send after withdrawing from \TC. Then, we tag the depositors and withdrawers in distance~2 using manually crawled \href{https://etherscan.io/accounts/label/}{labels from Etherscan}. We finally cluster the addresses into different platforms based on their labels.

 \begin{figure}[t]
\centering
    \includegraphics[width=\columnwidth]{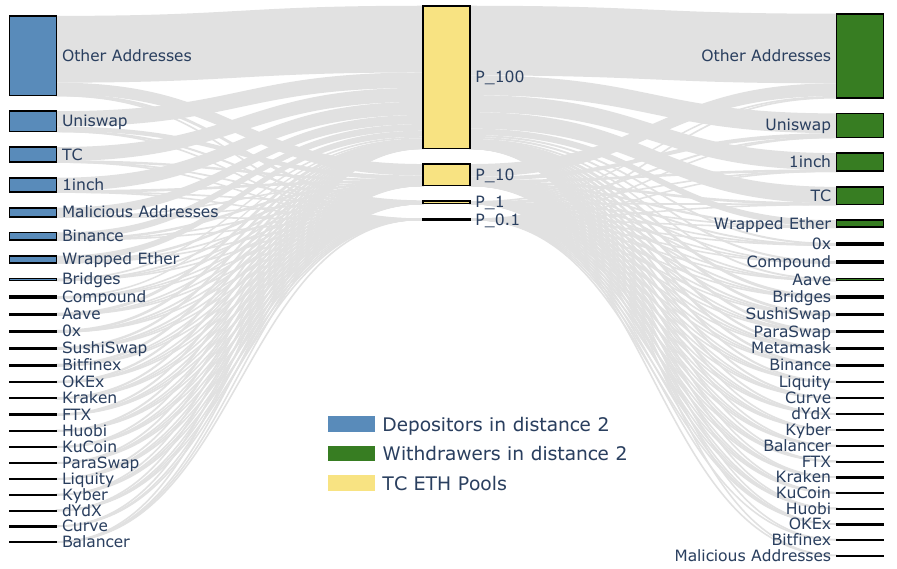}
  \caption{Prior-sanction \TC \ETH pools coin flow. The shown bandwidth of each flow represents the magnitude of the aggregate \ETH transferred from depositors in distance 2 to \TC (via depositors in distance 1), or from \TC to withdrawers in distance 2 (via withdrawers in distance 1).}
  \Description{Prior-sanction \TC \ETH pools coin flow. The shown bandwidth of each flow represents the magnitude of the aggregate \ETH transferred from depositors in distance 2 to \TC (via depositors in distance 1), or from \TC to withdrawers in distance 2 (via withdrawers in distance 1).}
  \label{fig:moneyFlow}
\end{figure}

Fig.~\ref{fig:moneyFlow} visualizes the \ETH flow via four \TC \ETH pools before \sanctiondate. We observe that the top 10 clusters in distance~$2$ cover {$48.11\%$} of the total deposit volume, and transfers from {\DEXes, e.g., Uniswap, } alone amount to {$750.2$K} \ETH ({$21.74\%$} of the total deposit volume). \DEXes are also the most popular \DeFi platforms to which \TC users transfer their withdrawn \ETH ({$27.2\%$} of the total withdrawal volume). This is probably because users are swapping \ETH to other tokens on \DEXes.
We also observe that {$10.64\%$} of the total deposit volume is re-deposited into \TC. 




\subsection{Why Do Users Resort to Mixers?}

Based on the coin flow, we analyze mixer user behaviors and find the following motivations for adopting mixers.

\begin{figure}[t]
\centering
\includegraphics[width=\columnwidth]{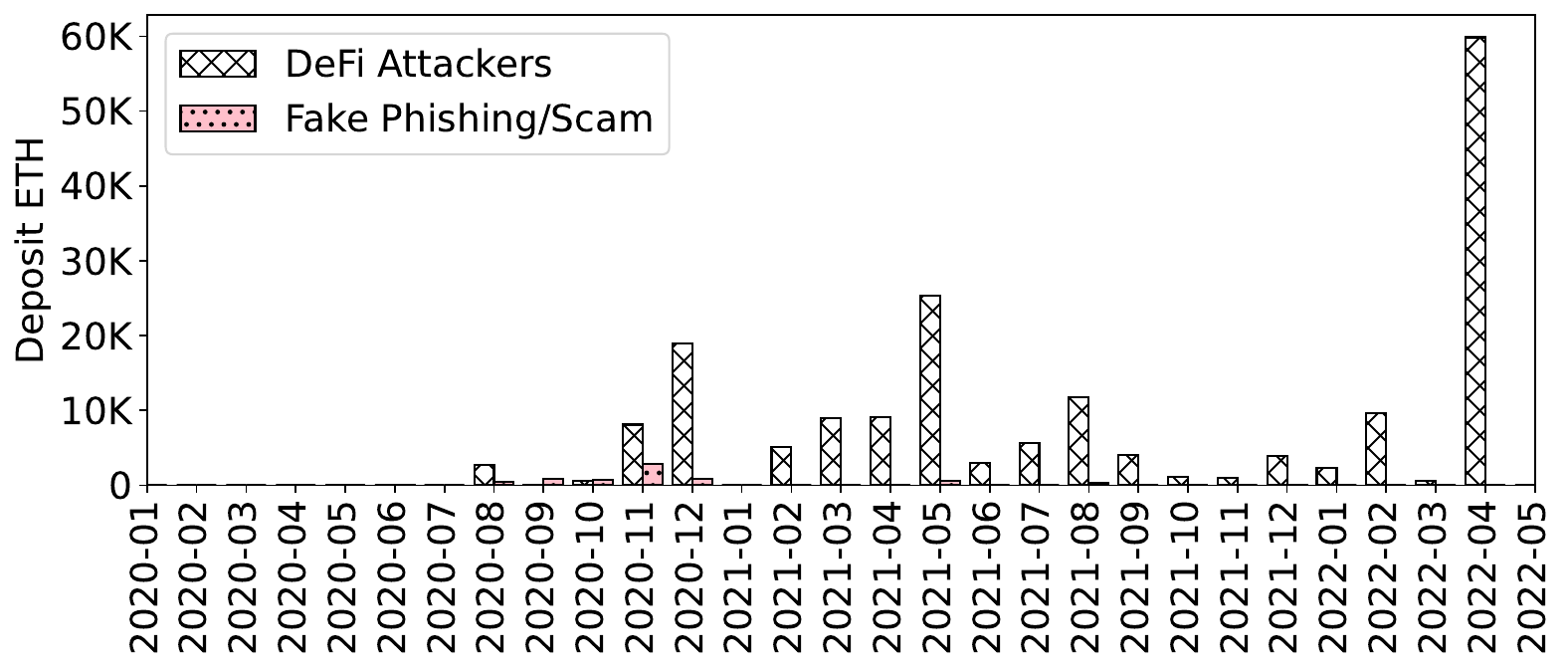}
\caption{TC deposit amounts of malicious addresses in distance 1 before May 2022. \DeFi Attackers deposit more assets than Fake Phishing/Scam addresses.}
\Description{TC deposit amounts of malicious addresses in distance 1 before May, 2022. \DeFi Attackers deposit more assets than Fake Phishing/Scam addresses.}
\label{fig:attackers_over_time}
\end{figure}

\begin{table}[htbp]
    \centering
     \caption{\numtotalmixerAttackers malicious addresses and \numtotalmixerBEVextractors \BEV extractors leverage \TC to hide their traces. An address can appear in more than one distance of depositor and withdrawer set.
    }
    \renewcommand\arraystretch{1.}
    \resizebox{\columnwidth}{!}{ 
    \begin{tabular}{c|lrrr}
        \toprule
       \multirow{2}{*}{Pattern}           & \multirow{2}{*}{Address Type}                                                                            & \multirow{2}{*}{Total}           
        &\multicolumn{2}{c}{Distance} \\
        \cline{4-5}
        ~ &~ &~ & $n=1$         & $n=2$         \\
        \midrule
        \multirow{3}*{$\mathsf{Mixer} \xrightarrow{n} \mathsf{Malicious}~\mathsf{Addresses}$}  & Fake Phishing Scam  &$19$   & $13$  &$8$       \\
         ~ & DeFi Attacker  &$121$ &$81$   &$49$     \\
         ~&CEX Attacker &$1$  &$0$    &$1$ \\
       
       \hline
       
        \multirow{3}*{$\mathsf{Malicious}~\mathsf{Addresses} \xrightarrow{n}\mathsf{Mixer}$}  & Fake Phishing/Scam                                            &$58$ & $26$  &$35$           \\
       ~ & DeFi Attacker  &$113$ & $88$  &$56$         \\
       & CEX Attacker &$1$ &$0$  &$1$     \\
       
        \bottomrule
        \bottomrule
        
        \multirow{3}*{$\mathsf{Mixer} \xrightarrow{n} \mathsf{BEV}~\mathsf{Extractors}$}  & Sandwich Attacker                                            &$431$ & $240$  &$230$         \\
       ~ &Arbitrageur  &$72$ &$48$ & $45$          \\
       & Liquidator  &$51$ &$27$ &$34$        \\
       \hline
       
        \multirow{3}*{$\mathsf{BEV}~\mathsf{Extractors} \xrightarrow{n}\mathsf{Mixer}$}  &Sandwich Attacker  &$2{,}185$  &$495$ &$2{,}096$          \\
         ~ &Arbitrageur  &$128$ &$33$   &$124$     \\
         ~&Liquidator &$73$  &$56$    &$60$ \\
        \bottomrule
        
    \end{tabular}
    }
   
    \label{tab:money_tracing_hackers_and_bev}
\end{table}

\smallskip\noindent\textbf{Money laundering:}
Because mixers break the linkability between addresses, users can use them  to conceal their traces. To do so, users withdraw \ETH from a mixer pool to a fresh address, and then transfer their assets (via intermediary addresses) to \CEXes, e.g., Binance and Huobi, to receive fiat currencies. We crawl 364 labeled \CEX addresses from Etherscan and identify that {$63$} out of them appear in the \TC withdrawer sets {in distance $2$}, \fixed{which may attempt to leverage intermediary addresses to hide their traces.} We find that {$4{,}062$} addresses transfer {$26.2$K~\ETH ($34.85$M USD)} into \CEXes.

\smallskip\noindent\textbf{Anonymity mining}:
\TC incentivizes users to adopt mixers through AM~\cite{TornadoCashGovernanceProposal}. Users can earn rewards for depositing and withdrawing funds from a \TC \ETH pool, and interacting with \TC anonymity mining contract (see Section~\ref{sec:incentivisedpools} for more details).
Our findings show that {$1{,}141$} depositors and {$1{,}290$} withdrawers are used to receive \AM rewards, while depositing {$532.3$K}~\ETH ({$707.98$M} USD) and withdrawing {$512.6$K}~\ETH ({$681.79$M} USD) respectively. Furthermore, we find that addresses using \AM typically deposit and withdraw multiple times. For instance, among the top $100$ withdrawers with the highest withdrawal amount, {$40$} addresses received \AM rewards.
     

\smallskip\noindent\textbf{Extracting \BEV:} Mixers also provide opportunities to \BEV extractors (a \emph{\BEV extractor} is an address which is used to perform a sandwich attack, liquidation, or arbitrage) to enhance their privacy. To understand how many \BEV extractors utilize \TC, we contacted the authors of~\cite{qin2022quantifying} to reuse their quantification results on sandwich attacks, liquidations, and arbitrage from block $6{,}803{,}256$ (December 1st, 2018) to block $12{,}965{,}000$ (August 5th, 2021). We then analyze whether the $11{,}289$ \BEV extractors identified in~\cite{qin2022quantifying} appear in \TC depositor and withdrawer sets. We find that {$2{,}185$} addresses are used for sandwich attacks, $128$ for arbitrages, and $73$ for liquidations (cf.~Table~\ref{tab:money_tracing_hackers_and_bev}), while depositing \totalDepositMixerBEV~\ETH ({$154.25$M USD}) into \TC. Furthermore, \numwithdrawBEVextractors \BEV extractors withdraw \totalWithdrawMixerBEV~\ETH from \TC.

\smallskip\noindent{\textbf{Launching attacks:} Malicious actors may adopt mixers to hide their identities. To gain initial insights into how malicious users adopt \TC, we first crawl $6{,}611$ blockchain phishing- and attack-related addresses from the dataset provided by the DeFi Attack SoK~\cite{zhou2022sok}. This dataset contains data from \emph{(i)} Etherscan, \emph{(ii)} Rekt News, \emph{(iii)} Slowmist, \emph{(iv)} Cryptosec, and \textit{(v)} CryptoscamDB. We regard the $6{,}611$ addresses as \emph{malicious addresses} and find that \numtotalmixerAttackers addresses out of them appear in \TC depositor and withdrawer sets.

We find that \numdepositAttackers malicious addresses deposit \totalDepositMixerAttackers~\ETH ({$258.62$M} \USD) into \TC, while \numwithdrawAttackers addresses withdraw \totalWithdrawMixerAttackers~\ETH from \TC (cf.~Table~\ref{tab:money_tracing_hackers_and_bev}). We further cluster the \numtotalmixerAttackers malicious addresses into three categories: \emph{(i)} Fake Phishing/Scam {($31.22\%$)}, which are labeled as ``Phish / Hack'' on Etherscan or scam addresses on CryptoscamDB; \emph{(ii)} \DeFi attackers {($68.29\%$)}, which attacked a \DeFi platform; \emph{(iii)} \CEX attackers {($0.49\%$)}, which steal assets from a \CEX.}

Fig.~\ref{fig:attackers_over_time} shows the malicious addresses directly depositing \ETH into \TC overtime. Malicious addresses seem to be careful to use mixers: The first time a malicious address deposits \ETH into \TC is in July, 2020, when the anonymity set size exceeds $1{,}000$ (cf. Fig.~\ref{fig:mixer_depositor_over_time}).

\begin{figure}
     \centering
  \includegraphics[width=\columnwidth]{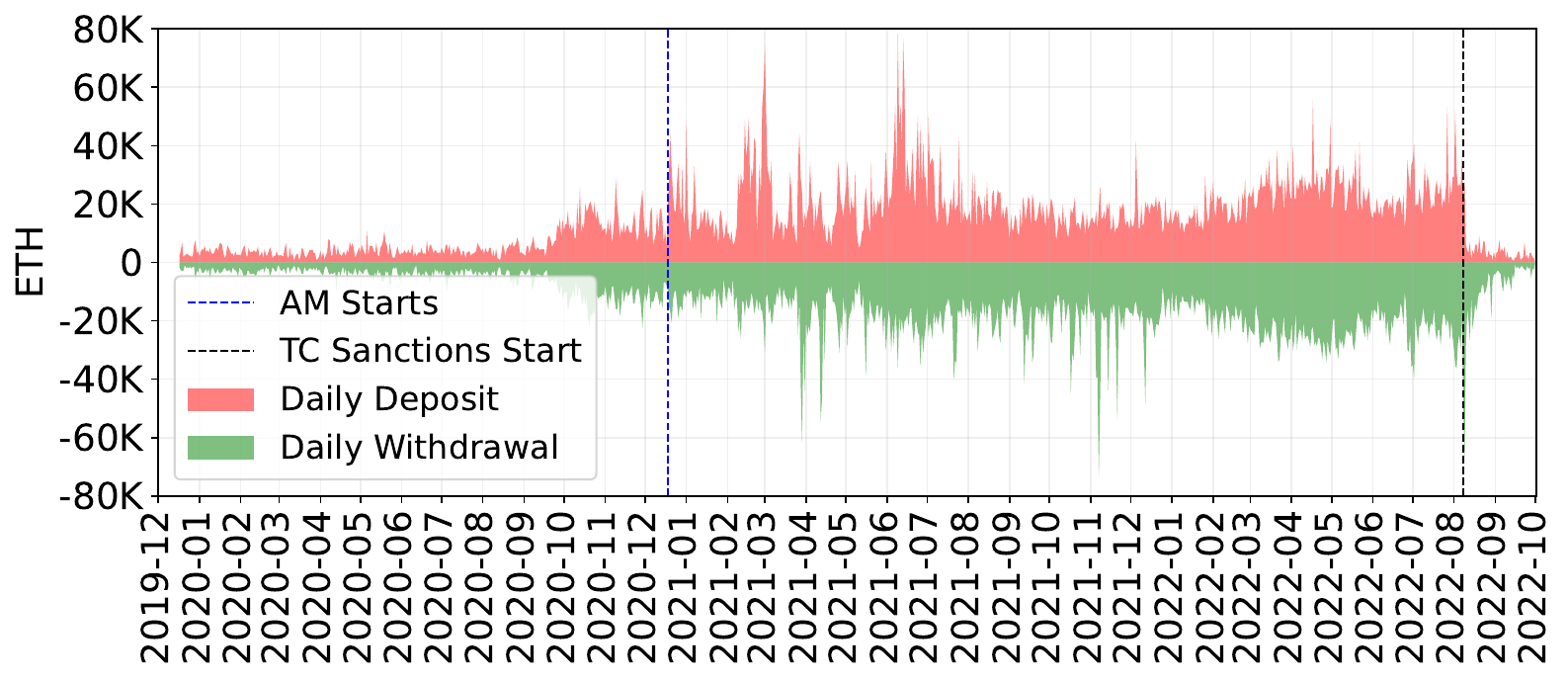}
\caption{Daily transactions in \TC \ETH pools. There was a panic exit when the \OFAC sanctions were announced.}
\Description{Daily transactions in \TC \ETH pools. There was a panic exit when the \OFAC sanctions were announced.}
\label{fig:pool_daily_tc}
\end{figure}


\begin{figure}[t]
   \centering
  \includegraphics[width=\columnwidth]{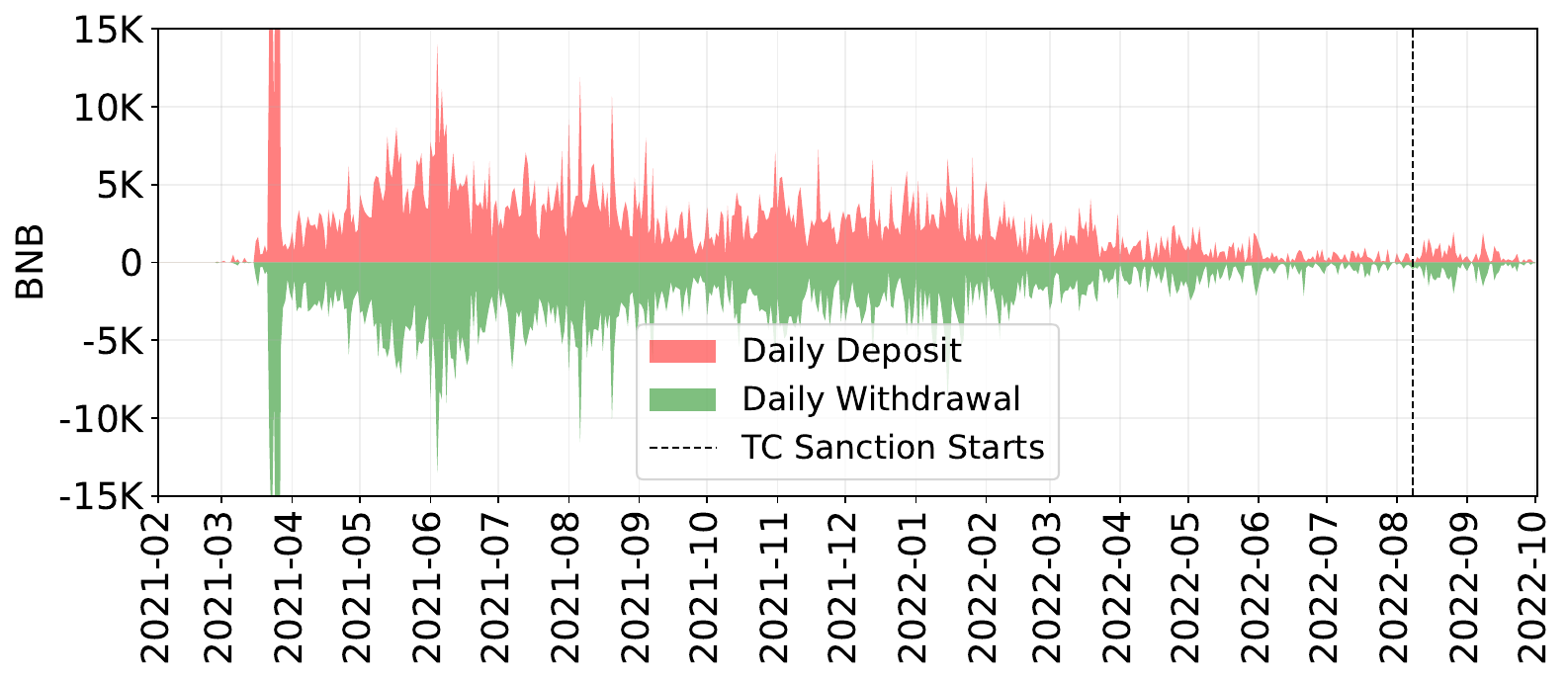}
\caption{Daily transactions in \TN \BNB pools. There were almost zero deposits and withdrawals in \TN during June and July 2022, but the activities increased after the \TC sanctions.}
\Description{Daily transactions in \TN \BNB pools. There were almost zero deposits and withdrawals in \TN during June and July 2022, but the activities increased after the \TC sanctions.}
\label{fig:pool_daily_tn}
\end{figure}


 \begin{figure}[t]
\centering
    \includegraphics[width=\columnwidth]{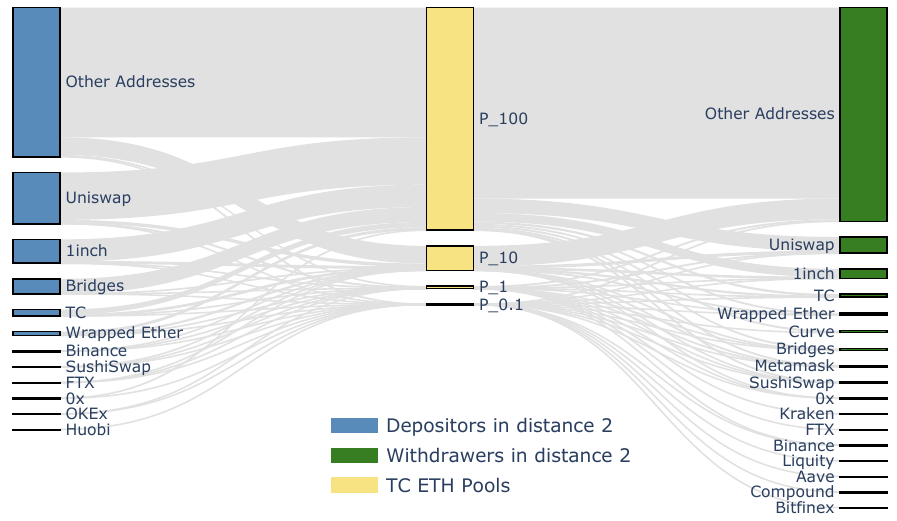}
  \caption{Post-sanction \TC \ETH pools coin flow. After the sanctions, more than $85\%$ of the \TC withdrawn \ETH are transferred to intermediary addresses in distance $2$, rather than being transferred to \DeFi platforms or \CEXes.}
  \Description{Post-sanction \TC \ETH pools coin flow. After the sanctions, more than $85\%$ of the \TC withdrawn \ETH are transferred to intermediary addresses in distance $2$, rather than being transferred to \DeFi platforms or \CEXes.}
  \label{fig:post_sanction_coin_flow}
\end{figure}

\begin{figure}[t]
\centering
  \includegraphics[width=\columnwidth]{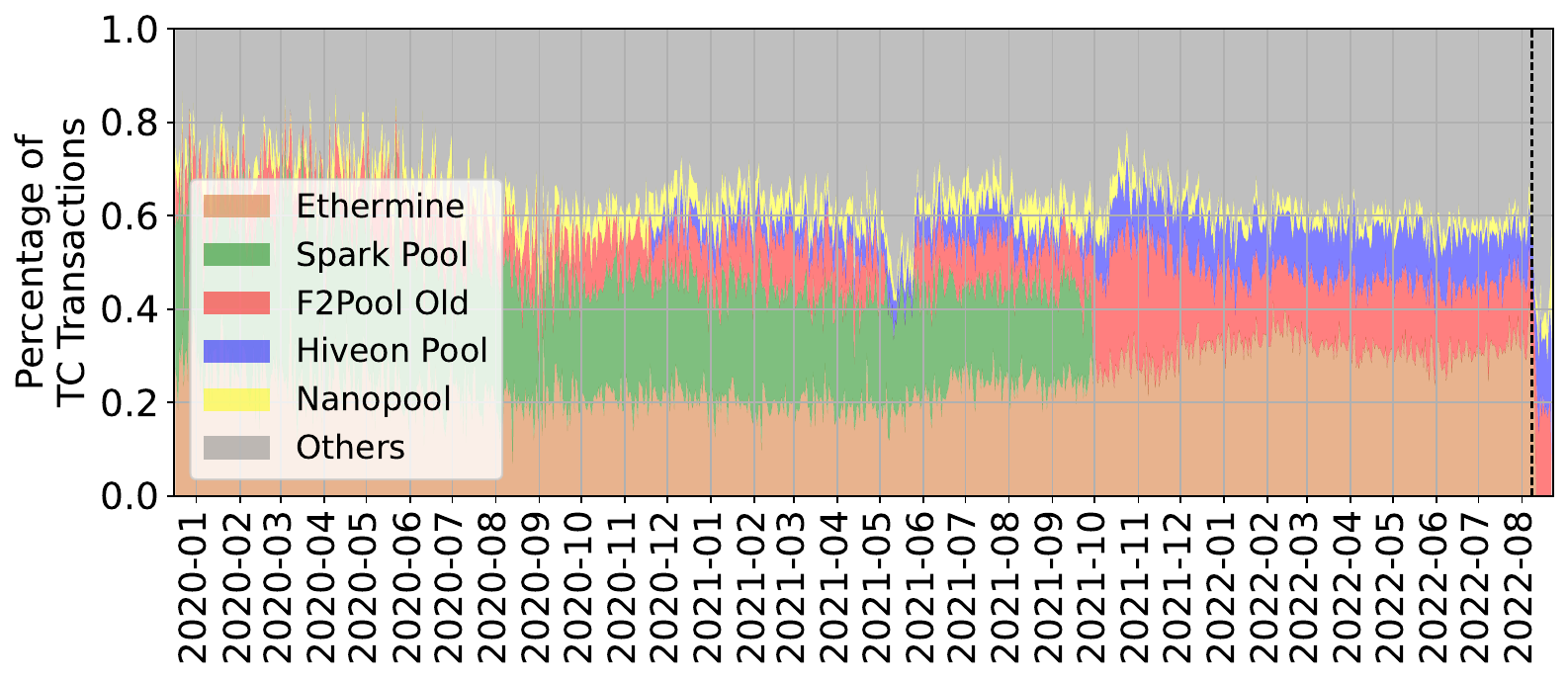}
\caption{\TC transactions mined by various mining pools before August 23rd, 2022. \href{https://etherscan.io/address/0xEA674fdDe714fd979de3EdF0F56AA9716B898ec8}{Ethermine} mined the most \TC transactions before \sanctiondate, but stopped processing TC transactions after the sanctions are announced.}
\label{fig:pool_daily_tc_mined}
\end{figure}

\section{\OFAC Sanctions Impact on \ZKP Mixers}\label{sec:OFAC-sanctions}

In this section, we investigate how \OFAC sanctions affect mixers. 



\smallskip\noindent\textbf{Impact on Mixer Usage.}
To understand how users interact with \ZKP mixers before and after the sanctions, we plot the daily deposited and withdrawn \ETH and \BNB in \TC and \TN pools from \crawlstartdate to \crawlenddate in Figures~\ref{fig:pool_daily_tc} and~\ref{fig:pool_daily_tn}. We observe that the graphs of daily deposits and withdrawals seem to be approximately symmetrical before the TC sanctions were announced (i.e., \sanctiondate). Interestingly, there was a panic exit on \sanctiondate: $230$ \TC withdrawers withdrew their $48{,}900$~\ETH due to the sanctions. The \TC daily deposits decreased by approximately {$83\%$} after \sanctiondate. Moreover, there were almost zero daily deposits and withdrawals in \TN during July 2022, but there was a tiny increase in August after the sanctions (cf.~Fig.~\ref{fig:pool_daily_tn}). This is likely because privacy-seeking users leverage \TN to replace \TC to hide their identities.

\smallskip\noindent\textbf{Post-Sanction \TC Deposits.} Although the \TC official websites are banned by the US \OFAC, users can still interact with \TC contracts (e.g., through \href{https://github.com/tornadocash/tornado-cli}{\TC \CLI}) to deposit and withdraw assets. We notice that the deposits in \TC are not zero after the sanctions started: from block \tcSanctionStartBlock (August 9th, 2022) to \crawlendEthdBlock (\crawlenddate), $487$ addresses deposited $47{,}056.8$~\ETH~($62.59$M~USD) into \TC pools. Only $75$ ($15.40\%$) out of the $487$ addresses ever deposited \TC before the sanctions started.

\smallskip\noindent\textbf{Post-Sanction \TC Coin Flow.} Moreover, we find that $671$ addresses withdraw {$170{,}826.3$~\ETH ($227.20$M~USD)} from \TC \ETH pools. To understand the post-sanction \TC \ETH pools coin flow, we adopt extend the mixer pools to cover the distance $2$ depositors and withdrawers. As shown in Fig.~\ref{fig:post_sanction_coin_flow}, we observe that after \sanctiondate, more than {$85.49\%$} of the withdrawn \ETH are transferred to intermediary addresses in distance 2, before interacting with \CEXes or \DeFi platforms. We speculate this is likely because \TC users attempt to bypass the censorship of \CEXes or \DeFi platforms, which claim to ban addresses receiving assets from \TC~\cite{chainalysis-tc-sanctions-challenges}.






\smallskip\noindent\textbf{Impact on Mining \TC Transactions.} \OFAC sanctions against \TC also have an influence on Ethereum miners. As shown in Fig.~\ref{fig:pool_daily_tc_mined}, we plot the distribution of \TC transactions mined by various mining pools over time. \href{https://etherscan.io/address/0xEA674fdDe714fd979de3EdF0F56AA9716B898ec8}{Ethermine} is the largest mining pool that mined the most \TC transactions before \sanctiondate. However, we observe that, after the sanctions started, Ethermine stopped processing any transactions related to deposits and withdrawals in \TC (cf.~Fig.~\ref{fig:pool_daily_tc_mined}).

\section{Incentivized ZKP Mixer Pools}\label{sec:incentivisedpools}
Spearheaded by the introduction of AMR~\cite{le2020amr}, we have witnessed a number of real-world mixer pools~\cite{TornadoCashGovernanceProposal} (cf.\ Section~\ref{subsec:mixing-service}) introducing rewarding governance tokens through \emph{anonymity mining} (\AM). In this section, we analyze how \AM affects user privacy.

\subsection{Anonymity Mining in TC Pools}
\begin{figure}[t]
\centering
  \includegraphics[width=\linewidth]{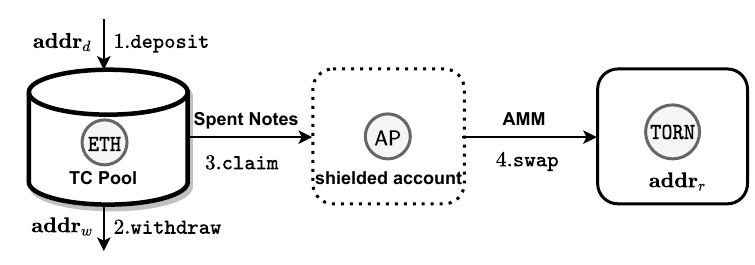}
  \caption{Overview of the \TC anonymity mining.}
  \Description{Overview of the \TC anonymity mining.}
  \label{fig:anonymity-mining}
\end{figure}

\TC incentivizes users to maintain their assets in \TC \ETH pools through \AM~\cite{TornadoCashGovernanceProposal}. Users receive \TORN tokens as rewards through a so-called shielded liquidity mining protocol as follows (cf.\ Fig.~\ref{fig:anonymity-mining}).


\noindent\emph{(1)} \texttt{Deposit}: A user deposits \ETH into a \TC pool using addresses $\mathbf{addr}_d$, and receives a deposit \emph{note}.

\noindent\emph{(2)} \texttt{Withdraw}: When the user withdraws \ETH from a \TC pool, the deposit note becomes a \emph{spent note}.

\noindent\emph{(3)} \texttt{Claim}: After withdrawing from a  pool, the user submits the \emph{spent note} to the pool to claim the Anonymity Points $\mathsf{AP}$. Because $\mathsf{AP}$ is determined by the deposit amount and duration (both are private information), $\mathsf{AP}$ is stored privately on a shielded account\footnote{According to \cite{TornadoCashGovernanceProposal}, a shielded account is a secret key newly generated by a user, which is used to encrypt and submit claim and withdrawal data without revealing the user’s identity. For recoverability, the user encrypts this secret key using his \ETH public key and stores the encrypted result on-chain.}.

\noindent\emph{(4)} \texttt{Swap}: A user can convert the shielded $\mathsf{AP}$ to public \TORN tokens using a dedicated \TC Automated Market Maker (AMM) exchange. The user receives the \TORN tokens in an address $\mathbf{addr}_r$ that can be different from the user's deposit or withdrawal address. 

\begin{equation}\label{eq:ap}
\centering
\mathsf{AP}_{\mathbf{u}}(t) = \sum_{p\in\{0{.}1, 1,10, 100\}}{ \mathsf{Weight}_p \cdot  \sum_{i=1}^{v_{p}}{\left(t^w_{p,i} - t^d_{p,i}\right)}}
\end{equation}

Equation~\ref{eq:ap} from \TC outlines the amount of $\mathsf{AP}$ a user $\mathbf{u}$ is entitled to at time $t$, where $\mathsf{Weight}_p$ is a predefined parameter to calculate a user's $\mathsf{AP}$ in various pools. {$\mathsf{Weight}_p$ is predefined as $10$, $20$, $50$ and $400$ in \TC $0.1$, $1$, $10$, and $100$ \ETH pools, respectively}. $v_{p}$ corresponds to the number of withdrawals in the $\mathbf{P}_p$ pool before time $t$. $t^d_{p,i}$ and $t^w_{p,i}$ are the block numbers of $\mathbf{u}$'s $i$-th deposit and withdrawal, $0 \le i \le v_{p}$. For instance, if a user $\mathbf{u}$ deposits twice $1$ \ETH into $\mathbf{P}_{1}$ at block $11{,}476{,}000$ and $11{,}476{,}100$, and deposits $10$ \ETH into the $\mathbf{P}_{10}$ pool at block $11{,}476{,}000$, and $\mathbf{u}$ withdraws all the deposited funds at block $11{,}476{,}200$, then $\mathbf{u}$'s $\mathsf{AP}$ is $20 \times (100+200) + 400 \times 200 = 86{,}000$.

\subsection{Linking User Addresses through \AM}
\label{sec:am-linking}
\AM aims to attract users to deposit more coins over a longer timeframe. However, \AM also increases the required user interactions with mixers (e.g., claiming to receive rewards), and may thus provoke the leakage of privacy-compromising information. We explore how to link users' withdrawals and deposits by solving Equation~\ref{eq:ap}.

We first identify the addresses that received \TORN tokens from \TC pools. From block $11{,}474{,}710$ (\amstartdate) to \crawlendEthdBlock (\crawlenddate), we identify {$15{,}659$} \href{https://etherscan.io/address/0x5cab7692d4e94096462119ab7bf57319726eed2a\#tokentxns}{TC Reward Swap} events, and find that {$1{,}844$} addresses received \TORN. We then extract the converted $\mathsf{AP}$ value in swap events. 



\smallskip\noindent \textbf{Receive Rewards with Deposit Address.}
In the following, we show that re-using a deposit address to receive rewards can deteriorate a user's privacy. We discover that among the {$1{,}844$} addresses receiving \TORN, {$1{,}141$} are depositors. We extract their deposit time, receiving \TORN time, and the converted values of $\mathsf{AP}$. Based on the data, we divide the {$1{,}141$} depositors into three categories:
\begin{itemize}


    \item \emph{$1$ deposit/$1$ claim/$1$ pool:} Out of the {$1{,}141$} depositors, {$236$} only deposited \emph{once} in \emph{one} TC pool and only received \TORN tokens from $\mathsf{AP}$ with \emph{one} transaction. In this case, Equation~\ref{eq:ap} can be simplified as $\mathsf{AP}_{\mathbf{u}}(t) = \mathsf{Weight}_p \cdot (t^w_{p,1} - t^d_{p,1})$. Because $\mathsf{AP}_{\mathbf{u}}(t)$ and $t^d_{p,1}$ are known, we can resolve the value of $t^w_{p,1}$ and search if there is a withdrawal transaction in block $t^w_{p,1}$. In total, we find the withdrawals for {$53$} depositors. For the remaining depositors, we speculate that they have likely not yet converted all their $\mathsf{AP}$.

    \item \emph{$n$ deposits/$1$ claim/$1$ pool:} {193} addresses deposited \emph{more than once} in \emph{one} \TC pool but only received \TORN \emph{once}. Equation~\ref{eq:ap} can be simplified as $\mathsf{AP}_{\mathbf{u}}(t) = \mathsf{Weight}_p \cdot \sum_{i=1}^{v_{p}}{(t^w_{p,i} - t^d_{p,i})}$. In this case, we find the possible withdrawals for {$51$} depositors.
    
    \item \emph{$n$ deposits/$n$ claims/$n$ pools:}  For the remaining depositors receiving \TORN \emph{more than once} or using \emph{multiple} pools, it is challenging to find their withdrawals, because it is uncertain whether they have claimed all $\mathsf{AP}$ and Equation~\ref{eq:ap} is hard to solve. However, we would suggest users avoid reusing addresses to receive \TORN, because one conversion of $\mathsf{AP}$ for a depositor shows that this depositor has already (partly or entirely) withdrawn the deposits. 
\end{itemize}

In total, we can find the possible withdrawal transactions for {$104$} addresses, indicating that re-using a deposit address for receiving AM rewards can deteriorate users' privacy.







\begin{figure}[t]
\centering
  \includegraphics[width=\linewidth]{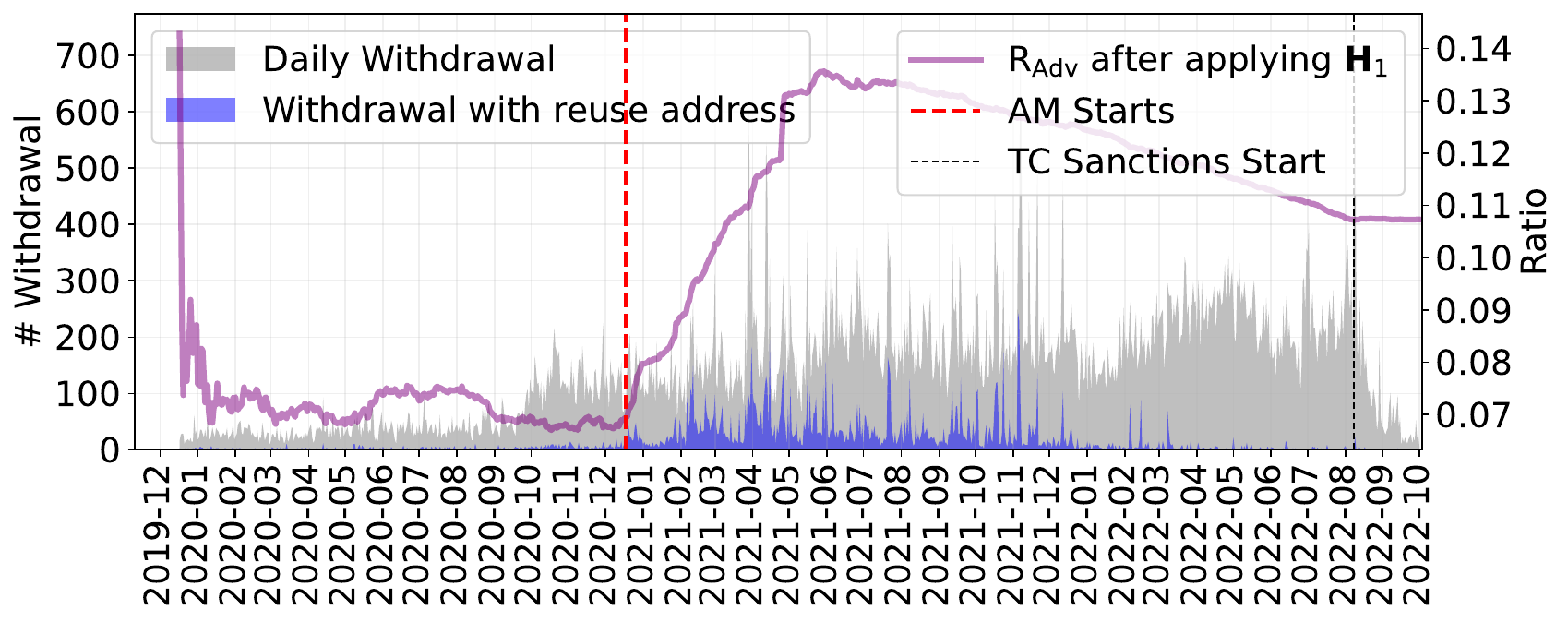}
  \caption{The AM launch does not increase the number of daily withdrawals but attracts privacy-ignorant users. Heuristic 1 performs better after \AM started, i.e., the \increaseR of the probability that an adversary links a withdrawer to the correct depositor (cf. Eq.~\ref{eq:avg-increase}), rises from $7.00\%$ to $13.50\%$.}
  \Description{The AM launch does not increase the number of daily withdrawals but attracts privacy-ignorant users. Heuristic 1 performs better after \AM started, i.e., the \increaseR of the probability that an adversary links a withdrawer to the correct depositor (cf. Eq.~\ref{eq:avg-increase}), rises from $7.00\%$ to $13.50\%$.}
  \label{fig:TC_Pools_Daily_withdrawals}
\end{figure}

\subsection{\AM's Impact on Mixer Anonymity Set}\label{sec:am-as}

To understand how \AM affects a mixer pool's anonymity set, we investigate the privacy-ignorant addresses attracted by \AM.
As shown in Fig.~\ref{fig:TC_Pools_Daily_withdrawals}, we first plot the number of daily withdrawal transactions in \TC \ETH pools. We then highlight the withdrawals in which deposit addresses are reused to receive withdrawn assets.


We observe that the daily withdrawals in \TC \ETH pools are not affected by \AM as intended: the number started increasing before \AM launch on October 18th, 2020. However, \AM does attract more users who reuse the deposit addresses to withdraw. Such ``reusing depositors'' are likely interested in mining \TORN, but privacy-ignorant. 

Based on our observations, we introduce the following heuristic, which identifies privacy-ignorant users that reuse addresses. We apply Heuristic~1 to prune privacy-ignorant user addresses and compute a more accurate mixer anonymity set size (see~Section~\ref{sec:h1} for more details). We observe that Heuristic~$1$ performs better after \AM started. As shown in Fig.~\ref{fig:TC_Pools_Daily_withdrawals}, the \increaseR $\mathsf{R}_\mathsf{Adv}$ (cf. Eq.~\ref{eq:avg-increase}) that an adversary links a withdrawer to the correct depositor rises from $7.00\%$ (before \AM) to $13.50\%$ (after \AM) on average.

\begin{figure*}[t]
\centering
\hfill
\subfigure[Heuristic 1.]{
\includegraphics[width=0.34\columnwidth]{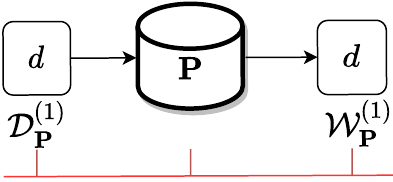}
\label{fig:overlap}
}%
\hfill
\subfigure[Heuristic 2.]{
\includegraphics[width=0.34\columnwidth]{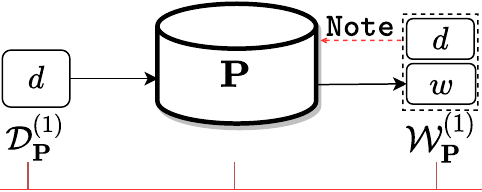}
\label{fig:heuristic_sender}
}%
\hfill
\subfigure[Heuristic 3.]{
\includegraphics[width=0.34\columnwidth]{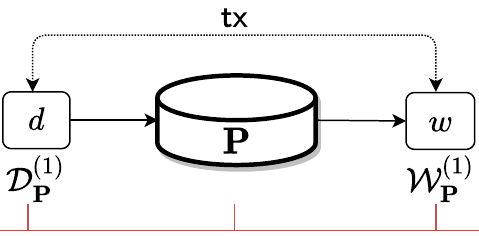}
\label{fig:related}
}%
\hfill
\subfigure[Heuristic 4.]{
\includegraphics[width=0.34\columnwidth]{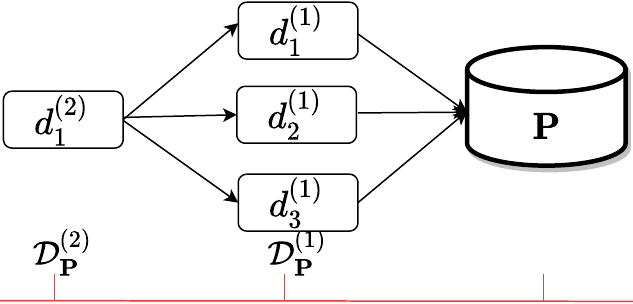}
\label{fig:bridge_address}
}%
\hfill
\subfigure[Heuristic 5.]{
\includegraphics[width=0.34\columnwidth]{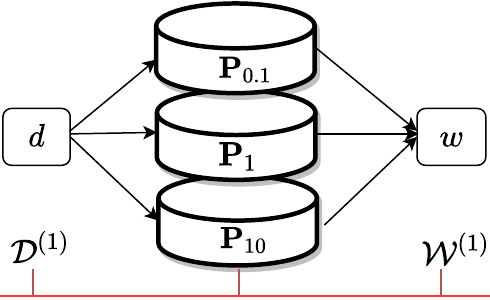}
\label{fig:h_cross_pool}
}%
\hfill
\vspace{-0.5cm}
\centering
\caption{{Overview of Heuristics 1-5:\protect\\
\emph{(a)} $\mathbf{H_1}$:  
A user applies the same address $d$ for deposit and withdrawal.\protect\\ 
\emph{(b)} $\mathbf{H_2}$:  A user adopts an address $w$ to receive the withdrawn \coins but a deposit address to pay the withdrawal transaction fees.\protect\\ 
\emph{(c)} $\mathbf{H_3}$:  A user adopts two distinct addresses $d$ and $w$ to deposit and withdraw in $\mathbf{P}$, while $d$ and $w$ are related in a transaction $\mathsf{tx}$. \protect\\ \emph{(d)} $\mathbf{H_4}$:  An address $d^{(2)}_{1}$ in distance 2  controls $3$ intermediary addresses $d^{(1)}_{j} (j = 1, 2 ,3)$ in the distance 1, to deposit \coins in $\mathbf{P}$. \protect\\ \emph{(e)} $\mathbf{H_5}$:  A user adopts an address $d$ to deposit  into $\mathbf{P}_{0.1}$, $\mathbf{P}_{1}$ and $\mathbf{P}_{10}$, and uses address $w$ to withdraw the same times from the pools.}}
\Description{{Overview of Heuristics 1-5:\protect\\
\emph{(a)} $\mathbf{H_1}$:  
A user applies the same address $d$ for deposit and withdrawal.\protect\\ 
\emph{(b)} $\mathbf{H_2}$:  A user adopts an address $w$ to receive the withdrawn \coins but a deposit address to pay the withdrawal transaction fees.\protect\\ 
\emph{(c)} $\mathbf{H_3}$:  A user adopts two distinct addresses $d$ and $w$ to deposit and withdraw in $\mathbf{P}$, while $d$ and $w$ are related in a transaction $\mathsf{tx}$. \protect\\ \emph{(d)} $\mathbf{H_4}$:  An address $d^{(2)}_{1}$ in distance 2  controls $3$ intermediary addresses $d^{(1)}_{j} (j = 1, 2 ,3)$ in the distance 1, to deposit \coins in $\mathbf{P}$. \protect\\ \emph{(e)} $\mathbf{H_5}$:  A user adopts an address $d$ to deposit  into $\mathbf{P}_{0.1}$, $\mathbf{P}_{1}$ and $\mathbf{P}_{10}$, and uses address $w$ to withdraw the same times from the pools.}}
\label{fig:heuristics}
\end{figure*}

\smallskip\noindent\textbf{Heuristic for Address Reuse} ($\mathbf{H_1}$). 
If an address appears both in the depositor and withdrawer sets, then the deposits and withdrawals of this address are conducted by the same user (cf.\ Fig.~\ref{fig:overlap}).

In conclusion, contrary to the claims of related work~\cite{le2020amr}, we find that \AM does not always contribute to the mixers' anonymity set size as expected, because it attracts privacy-ignorant users.





\begin{table*}[t]
\centering

\caption{\textbf{Heuristics 1-5} applied to \TC \ETH and \TN \BNB pools before \crawlenddate. $|\sas{n}|$ represents the Anonymity Set Size after applying Heuristic $n$. The percentages show the difference between the simplified anonymity set and $\mathbf{OAS}_{\mathbf{P}}(t)$.
}

\renewcommand\arraystretch{1.1}
\resizebox{\textwidth}{!}{
\begin{tabular}{lr|r|r|r|r|r||r|r|r|r}
\toprule


 \multirow{3}{*}{\bf Pool} &\multirow{3}{*}{\multirow{1}{*}{$|\observedAS|$} } 
 &\multirow{3}{*}{$|\sas{1}|$}
  &\multirow{3}{*}{$|\sas{2}|$}
   &\multirow{3}{*}{$|\sas{3}|$}
    &\multirow{3}{*}{$|\sas{4}|$}
     &\multirow{3}{*}{$|\sas{5}|$}
 &\multicolumn{4}{c}{\bf Heuristic Combinations}\\
 \cline{8-11}
~&
~&
~&
~&
~&
~&
& \multirow{2}{*}{$\mathbf{H_1} + \mathbf{H_2}$}
& \multirow{2}{*}{$\mathbf{H_1} + \mathbf{H_2} + \mathbf{H_3}$} 
& {$\mathbf{H_1} + \mathbf{H_2} $}
& {$\mathbf{H_1} + \mathbf{H_2} + \mathbf{H_3}$} \\

~&
~&
~&
~&
~&
~&
~&
~&
~& $+ \mathbf{H_3} + \mathbf{H_4}$
& $ + \mathbf{H_4} + \mathbf{H_5}$
\\
\hline


TC 0.1 \ETH
&{$11{,}941$}
& {$10{,}745 {\thinspace} (-10.02 \%)$}

 &{$11{,}894 {\thinspace} (-0.39 \%)$}

 &{$9{,}890 {\thinspace} (-17.18 \%)$}
 &{$11{,}439 {\thinspace} (-4.20 \%)$}
 &{$11{,}857 {\thinspace} (-0.70 \%)$} 

  &{$-10.38 \%$} &{$-26.76 \%$} &{$-30.07 \%$} &\cellcolor{Gray}{$-30.68 \%$}\\

TC 1 \ETH
&{$17{,}843$}  
& {$16{,}422 {\thinspace} (-7.96 \%)$} 
&{$17{,}791 {\thinspace} (-0.29 \%)$} 
&{$15{,}445 {\thinspace} (-13.44 \%)$}
&{$17{,}077 {\thinspace} (-4.29 \%)$} 
&{$17{,}733 {\thinspace} (-0.62 \%)$}
 &{$-8.28 \%$} &{$-20.83 \%$} &{$-24.44 \%$} &\cellcolor{Gray}{$-24.98 \%$}\\

TC 10 \ETH
&{$16{,}227$} 
& {$14{,}587 {\thinspace} (-10.11 \%)$}

&{$16{,}187 {\thinspace} (-0.25 \%)$} 

&{$14{,}348 {\thinspace} (-11.58 \%)$}
&{$15{,}460 {\thinspace} (-4.73 \%)$} 
&{$16{,}111 {\thinspace} (-0.71 \%)$} 
 &{$-10.38 \%$} &{$-20.34 \%$} &{$-24.51 \%$} &\cellcolor{Gray}{$-25.14 \%$}\\

TC 100 \ETH
&{$6{,}433$}  &{$5{,}608 {\thinspace} (-12.82 \%)$} 
& {$6{,}407 {\thinspace} (-0.40 \%)$}
 &{$5{,}754 {\thinspace} (-10.55 \%)$}
 &{$5{,}975 {\thinspace} (-7.12 \%)$} 
 &{$6{,}347 {\thinspace} (-1.34 \%)$}  &{$-13.23 \%$} &{$-21.14 \%$} &{$-27.47 \%$} &\cellcolor{Gray}{$-28.57 \%$}\\

\hline


TN 0.1 \BNB
&{$3{,}972$}   
&{$2{,}820 {\thinspace} (-29.00 \%)$}
&{$3{,}702 {\thinspace} (-6.80 \%)$} 

&{$2{,}501 {\thinspace} (-37.03 \%)$}
&{$3{,}946 {\thinspace} (-0.65 \%)$}
&{$3{,}934 {\thinspace} (-0.96 \%)$} 
&{$-33.91 \%$} &{$-54.00 \%$} &{$-54.76 \%$} &\cellcolor{Gray}{$-55.39 \%$}\\

TN 1 \BNB
&{$3{,}890$}   
&{$2{,}868 {\thinspace} (-26.27 \%)$}
&{$3{,}626 {\thinspace} (-6.79 \%)$}

&{$2{,}531 {\thinspace} (-34.94 \%)$}
&{$3{,}852 {\thinspace} (-0.98 \%)$} 
&{$3{,}845 {\thinspace} (-1.16 \%)$}
 &{$-29.23 \%$} &{$-48.15 \%$} &{$-49.56 \%$} &\cellcolor{Gray}{$-50.41 \%$}\\

TN 10 \BNB
&{$1{,}675$}  
&{$1{,}156 {\thinspace} (-30.99 \%)$}

&{$1{,}605 {\thinspace} (-4.18 \%)$} 

&{$1{,}068 {\thinspace} (-36.24 \%)$}
&{$1{,}666 {\thinspace} (-0.54 \%)$}
&{$1{,}635 {\thinspace} (-2.39 \%)$}

 &{$-33.07 \%$} &{$-47.58 \%$} &{$-48.48 \%$} &\cellcolor{Gray}{$-50.15 \%$}\\
TN 50 \BNB
&{$231$} &{$217 {\thinspace} (-6.06 \%)$} 

&{$217 {\thinspace} (-6.06 \%)$} 

&{$204 {\thinspace} (-11.69 \%)$}
&{$231 {\thinspace} (0.00 \%)$}
&{$213 {\thinspace} (-7.79 \%)$}
 &{$-12.12 \%$} &{$-21.65 \%$} &{$-21.65 \%$} &\cellcolor{Gray}{$-28.14 \%$}\\ 

\bottomrule
\end{tabular}
}
\label{tab: h1-3}
\end{table*}


\section{Measuring Mixer Anonymity Set Size}\label{sec:measuring-as}
In the following, we propose heuristics to measure a mixer pool's anonymity set size, which is more representative than the naive $\observedAS$. Our heuristics are best-effort methods and subject to known limitations~\cite{victor2020address, romiti2020cross}. We thus attempt to construct ground truth from side channels to validate our heuristics (cf.~Section~\ref{sec:heuristic-validation}).

\subsection{Linking Heuristics}\label{sec:heuristics_all}
We propose the following heuristics (cf.~Fig.~\ref{fig:heuristics}) to leverage on-chain data and insights from our empirical study to link addresses and prune the \observedAS. Table~\ref{tab:model-definitions} summarizes the extended system model and definitions which are used in our linking heuristics.

\subsubsection{$\mathbf{H_1}$ - Address Reuse}\label{sec:h1}

\smallskip\noindent\textbf{Observation:} 
We observe that an address can be reused to both deposit and withdraw, which could be incautious behavior and leak privacy~\cite{beres2021blockchain,victor2020address}.

\smallskip\noindent\textbf{Heuristic~1:}  
If an address appears both in the depositor and  withdrawer sets, we assume that the deposits and withdrawals of this address are conducted by the same user (cf.\ Fig.~\ref{fig:overlap}). We apply Eq.~\ref{eq:balance} in Table~\ref{tab:model-definitions} to compute a depositor's balance and extract the depositors with a positive balance to evaluate the anonymity set: $\sas{1} = \{\addr \mid \addr \in  \mathcal{D}_{\mathbf{P}}(t) \land \mathbf{bal}_\addr(t) > 0\}$.

\subsubsection{$\mathbf{H_2}$ - Improper Withdrawal Sender}\label{sec:sender}

\smallskip\noindent\textbf{Observation:} 
Incautious users may adopt a deposit address $\addr_w$ to receive the withdrawn funds, while paying the transaction fees using their deposit address $\addr_d$. This action infers that $\addr_d$ and $\addr_w$ are likely controlled by the same user. \fixed{This action might happen when users are not familiar with the mixer functionality, which can leak users' privacy.} 

\smallskip\noindent\textbf{Heuristic~2:} 
We assume that given a depositor-withdrawer pair $(\addr_d, \addr_w)$ in a pool, where $\addr_d$ is not a relayer\footnote{Relayers are addresses who help users withdraw coins from a mixer towards a new address by paying for the transaction fees, in exchange receive a share of the withdrawn coins.}, if $\addr_d$ generates a withdrawal and assigns $\addr_w$ to receive the withdrawn coins, then $\addr_d$ and $\addr_w$ belong to the same user (cf.\ Fig.~\ref{fig:heuristic_sender}), i.e., $\link(\addr_d, \addr_w) = 1$.

Let $\mathcal{S}^{\texttt{nt}}_{\mathbf{P}}(t)$ be the set of linked address pairs in a pool $\mathbf{P}$. Given $\mathcal{S}^{\texttt{nt}}_{\mathbf{P}}(t)$, we merge the balance of the linked addresses to simplify the pool state, and then compute the anonymity set: $\sas{2} = \{ \addr \mid \mathbf{bal}_\addr(t) > 0 ~ \land (\addr, \mathbf{bal}_\addr(t)) \in \simplified(\mathbb{S}_\mathbf{P}(t),\mathcal{S}^{\texttt{nt}}_{\mathbf{P}}(t))\}$.

\subsubsection{$\mathbf{H_3}$ - Related Deposit-Withdrawal Address Pair}

\smallskip\noindent\textbf{Observation:} 
To withdraw coins, users are encouraged to choose a new address with no links to the deposit address. However, we observe that, users may adopt different deposit and withdrawal addresses, which are directly linked through a coin transfer.

\smallskip\noindent\textbf{Heuristic~3:}
We assume that, given two addresses $\addr_d \in \mathcal{D}_{\mathbf{P}}(t)$ and $\addr_w \in \mathcal{W}_{\mathbf{P}}(t)$, if $\addr_d$  transferred (received) coins or tokens to (from) $\addr_w$ before time $t$, then $\addr_d$ and $\addr_w$ are related and under the control of the same user (cf.\ Fig.~\ref{fig:related}), i.e., $\link(\addr_d, \addr_w) = 1$. Let $\mathcal{S}^{\texttt{tx}}_{\mathbf{P}}(t)$ be the set of related depositor-withdrawer pairs in a pool $\mathbf{P}$. We simplify the pool state and compute the anonymity set: $\sas{3} = \{\addr \mid \mathbf{bal}_\addr(t) > 0 ~\land (\addr, \mathbf{bal}_\addr(t)) \in \simplify(\mathbb{S}_\mathbf{P}(t),\mathcal{S}^{\texttt{tx}}_{\mathbf{P}}(t))\}$.

\subsubsection{$\mathbf{H_4}$ - Intermediary Deposit Address}

\smallskip\noindent\textbf{Observation:} We observe that there are multiple depositors in distance 1 whose coins are all transferred from the same depositor in distance 2. Hence, these depositors in distance 1 are likely temporary addresses and are only used to transfer funds into a~mixer.

\smallskip\noindent\textbf{Heuristic~4:}
We hence assume that given two addresses $d^{(1)} \in \mathcal{D}^{(1)}_{\mathbf{P}}(t)$ and  $d^{(2)} \in \mathcal{D}^{(2)}_{\mathbf{P}}(t)$, if all $d^{(1)}$'s coins are transferred from $d^{(2)}$ and $d^{(2)}$ is a user account, then $\link(d^{(1)}, d^{(2)}) = 1$.


We denote $d^{(1)}$ as an \emph{intermediary deposit address}, $\mathcal{B}^{(1)}_{\mathbf{P}}(t)$ as the set of intermediary deposit address, and $\mathcal{B}^{(2)}_{\mathbf{P}}(t)$ as the set of user accounts in distance $2$ who transfer coins to an address in $\mathcal{B}^{(1)}_{\mathbf{P}}(t)$. For each  address $d^{(1)}$ in $\mathcal{B}^{(1)}_{\mathbf{P}}(t)$, we replace it by the address in $\mathcal{B}^{(2)}_{\mathbf{P}}(t)$ which transfers coins to $d^{(1)}$. We then compute: $\sas{4} = \{ \addr \mid \mathbf{bal}_\addr(t) > 0 ~\land \addr \in  \mathcal{B}^{(2)}_{\mathbf{P}}(t) \cup \mathcal{D}^{(1)}_{\mathbf{P}}(t) \setminus \mathcal{B}^{(1)}_{\mathbf{P}}(t)\}$.

\subsubsection{$\mathbf{H_5}$ - Cross-pool Deposit}

\smallskip\noindent\textbf{Observation:}
Current mixer pools only support the deposit and withdrawal of a \emph{fixed} coin denomination. When a user aims to mix an \emph{arbitrary} amount of coins, the user needs to interact with multiple pools and may not change the respective deposit (or withdrawal) address (cf.\ Fig.~\ref{fig:h_cross_pool}).

\smallskip\noindent\textbf{Heuristic~5}:
Given a depositor-withdrawer pair $(\addr_d, \addr_w)$, we assume $\link(\addr_d, \addr_w) = 1$ if: \emph{(i)} $\addr_d$ and $\addr_w$ are both in $m (m > 1)$ pools, \emph{(ii)} in each pool, $\addr_d$'s total deposit amount equals $\addr_w$'s withdrawal amount, and \emph{(iii)} for each $\addr_w$'s withdrawal transaction $\directtx^w$, at least one of $\addr_d$'s deposit transaction $\directtx^d$ is generated earlier than $\directtx^w$. 

Let $\mathcal{S}^{\texttt{cu}}$ be the set of address pairs $(\addr_d, \addr_w)$ that satisfy the above conditions. Given $\mathcal{S}^{\texttt{cu}}$, we simplify the state of a pool $\mathbf{P}$, and then compute the anonymity set $\sas{5} = \{\addr \mid \mathbf{bal}_\addr(t) > 0 ~\land (\addr, \mathbf{bal}_\addr(t)) \in \simplify(\mathbb{S}_\mathbf{P}(t),\mathcal{S}^{\texttt{cu}}_{\mathbf{P}}(t)) \}$.

\subsubsection{Linking and Measuring Results}
\begin{figure}[t]
    \centering
    \includegraphics[width=\columnwidth]{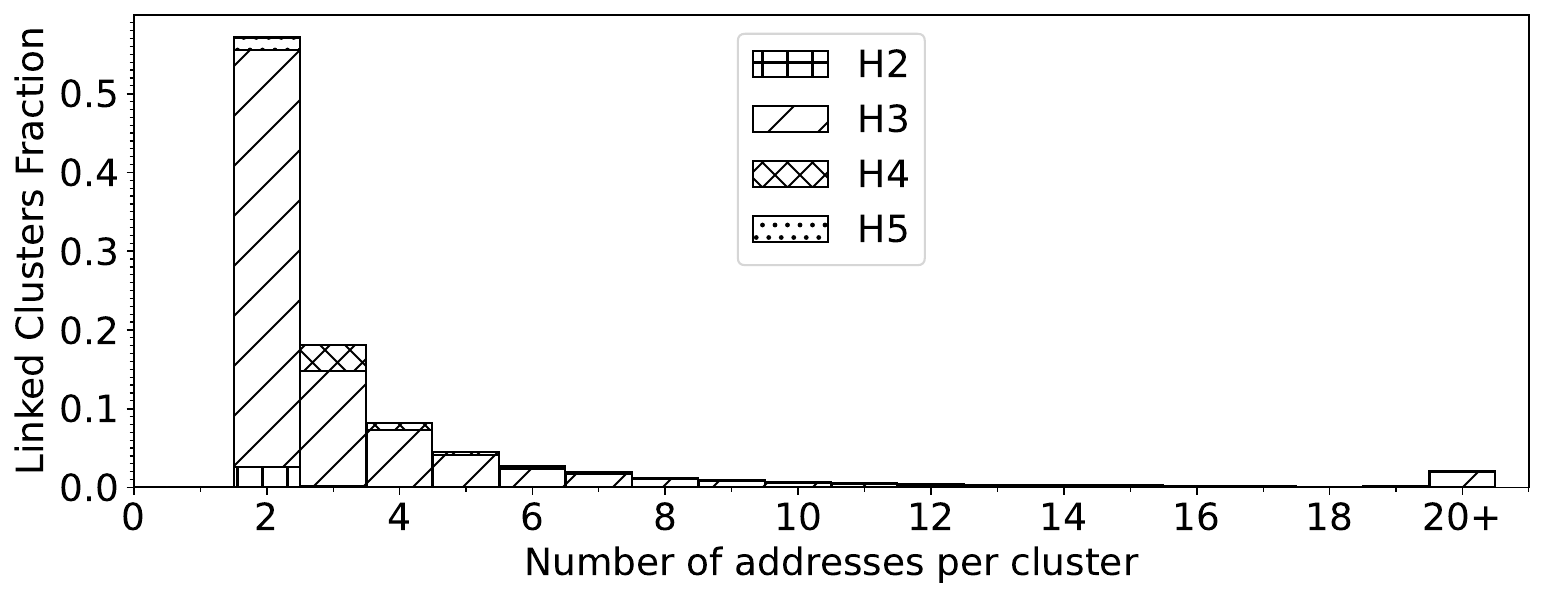}
    \caption{Number of addresses per \TC cluster. {$4{,}871\thinspace(57.14\%)$} \TC clusters only have two addresses.} 
    \Description{Number of addresses per \TC cluster. {$4{,}871\thinspace(57.14\%)$} \TC clusters only have two addresses.} 
    \label{fig:cluster_pdf_tc}
\end{figure}

Through Heuristics 2--5, we can link {$18{,}705$} \TC and {$9{,}383$} \TN address pairs, which form {$8{,}164$} and {$2{,}046$} clusters, respectively. Moreover, {$4{,}871\thinspace(57.14\%)$} \TC and {$1{,}190\thinspace (48.89\%)$} \TN clusters only have two addresses. Fig.~\ref{fig:cluster_pdf_tc} visualizes the distribution of \TC clusters over the number of addresses. Interestingly, we find that the cluster distribution is similar to previous works on Bitcoin address clustering (e.g., Fig.~9(b) in~\cite{karame2015misbehavior}).

Table~\ref{tab: h1-3} shows the \simplifiedAS of mixer pools after applying each heuristic individually. On \TC pools, Heuristic 1 reduces the anonymity set by an average of {$10.23\%$} from the reported \observedAS. For instance, in the \TC $100$ \ETH pool $\mathbf{P}_{100}$, there are {$6{,}433$} unique depositors, but only {$5{,}608$} depositors have a positive balance, and therefore contribute to the anonymity set. Consequently, \sas{1} of $\mathbf{P}_{100}$ is {$12.82\%$} less than the respective \observedAS.
For \TN, \sas{1} is reduced by an average of {$23.08\%$} from \observedAS.

 We can further reduce the \observedAS by combining two or more heuristics (cf. Table~\ref{tab: h1-3}). Combining all heuristics yields the largest reduction of \observedAS: after applying Heuristics 1-5 to the \TC (\TN) pools, an adversary can reduce the reported \observedAS on average by \totalTCASReduced~(\totalTNASReduced). Therefore, the probability that an adversary links a withdrawer (who withdraws at time $t$) to the correct depositor rises by ${37.63\% \thinspace (85.26\%)}$ on average (cf.~Eq.~\ref{eq:R_res}).
 
 \begin{equation}\label{eq:R_res}
    \mathsf{R}_\mathsf{Adv} = \frac{1/\left|\simplifiedAS\right|- 1/\left|\observedAS\right|}{1/\left|\observedAS\right|}={37.63\% \thinspace (85.26\%)} 
 \end{equation}




\subsubsection{User Privacy Behavior}
Our heuristics appear to function better on the \BSChain mixer (\TN) than on the ETH mixer (\TC). While our study should be repeated once the other mixers grow on both chains (e.g., Cyclone and \TP), our empirical evidence is the first to suggest a differing privacy-focus of users on ETH and \BSChain. One could also argue that privacy-aware users want the best available anonymity set, and will therefore use \TC and follow all best practices. As such, a suitable assumption is that anonymity set attracts anonymity set, i.e., the biggest anonymity set will inherently attract more users, and particularly those that worry about privacy (which is analogous to how liquidity attracts liquidity in financial exchanges).

\begin{table*}[t]
\centering
\caption{Validation attempt for TC linked address pairs. $\mathcal{S}^{\texttt{H}_i}_{\texttt{TC}}$ represents the linked address pairs obtained through Heuristic~$i$. For Heuristic 2, 3, and 5, Test Pairs =  depositors in $\mathcal{S}_{\texttt{GT}}$ (Candidate Ground Truth) $\times$  withdrawers in $\mathcal{S}_{\texttt{GT}}$. For Heuristic 4,  Test Pairs = distance-2 depositors in $\mathcal{S}_{\texttt{GT}}$ $\times$ distance-1 depositors in $\mathcal{S}_{\texttt{GT}}$.} 
\renewcommand\arraystretch{1.1}
\resizebox{0.9\textwidth}{!}{
\begin{tabular}{c|c|c|cccc|ccc}
\toprule
Candidate Ground Truth $\mathcal{S}_{\texttt{GT}}$
&Heuristics
&Test Pairs
&$tp=\mathcal{S}_{\texttt{GT}}\cap \mathcal{S}^{\texttt{H}_i}_{\texttt{TC}}$ &$tn=\overline{\mathcal{S}_{\texttt{GT}}}\cap \overline{\mathcal{S}^{\texttt{H}_i}_{\texttt{TC}}}$ &$fp= \overline{\mathcal{S}_{\texttt{GT}}}\cap \mathcal{S}^{\texttt{H}_i}_{\texttt{\texttt{TC}}} $ &$fn=\mathcal{S}_{\texttt{GT}}\cap \overline{\mathcal{S}^{\texttt{H}_i}_{\texttt{TC}}}$ &precision &recall &\bf F1\\
\hline

\multirow{5}*{$\mathcal{S}_{\texttt{Airdrop}}\thinspace(35{,}081)$}

&$\mathbf{H}_2$
&$ 931 \times 580 $
&$ 2 $  &$ 539{,}747 $ &$ 4 $ &$ 227 $ &$ 0.33 $ &$ 0.01 $ &\cellcolor{Gray}$ 0.02 $\\

~
&$\mathbf{H}_3$
&$ 931 \times 580 $
&$ 229 $  &$ 539{,}367 $ &$ 384 $ &$ 0 $ &$ 0.37 $ &$ 1.00 $ &\cellcolor{Gray}$ 0.54 $\\

~
&$\mathbf{H}_5$
&$ 931 \times 580 $
&$ 0 $  &$ 539{,}751 $ &$ 0 $ &$ 229 $ &$ 0.00 $ &$ 0.00 $ &\cellcolor{Gray}$ 0.00 $\\

\rowcolor{Gray}
\cellcolor{white}~
&$\mathbf{H}_2+\mathbf{H}_3+\mathbf{H}_5$
&$ 931 \times 580 $
&$ 229 $  &$ 539{,}366 $ &$ 385 $ &$ 0 $ &$ 0.37 $ &$ 1.00 $ &\cellcolor{Gray}$ 0.54 $\\

\cline{2-10}

~
&$\mathbf{H}_4$
&$ 710 \times 931 $
&$ 2 $  &$ 660{,}641 $ &$ 3 $ &$ 364 $ &$ 0.40 $ &$ 0.01 $ &\cellcolor{Gray}$ 0.01 $\\



\hline
\multirow{5}*{$\mathcal{S}_{\texttt{ENS}} \thinspace(5{,}105)$ }

&$\mathbf{H}_2$
&$ 291 \times 213 $
&$ 1 $  &$ 61{,}928 $ &$ 2 $ &$ 52 $ &$ 0.33 $ &$ 0.02 $ &\cellcolor{Gray}$ 0.04 $\\

~
&$\mathbf{H}_3$
&$ 291 \times 213 $
&$ 50 $  &$ 61{,}854 $ &$ 76 $ &$ 3 $ &$ 0.40 $ &$ 0.94 $ &\cellcolor{Gray}$ 0.56 $\\

~
&$\mathbf{H}_5$
&$ 291 \times 213 $
&$ 0 $  &$ 61{,}930 $ &$ 0 $ &$ 53 $ &$ 0.00 $ &$ 0.00 $ &\cellcolor{Gray}$ 0.00 $\\

\rowcolor{Gray}
\cellcolor{white}~
&$\mathbf{H}_2+\mathbf{H}_3+\mathbf{H}_5$
&$ 291 \times 213 $
&$ 50 $  &$ 61{,}854 $ &$ 76 $ &$ 3 $ &$ 0.40 $ &$ 0.94 $ &\cellcolor{Gray}$ 0.56 $\\

\cline{2-10}
~
&$\mathbf{H}_4$
&$ 118 \times 291 $
&$ 0 $  &$ 34{,}311 $ &$ 0 $ &$ 27 $ &$ 0.00 $ &$ 0.00 $ &\cellcolor{Gray}$ 0.00 $\\

\bottomrule
\end{tabular}
}
\label{tab:validation-attempt}
\end{table*}

\subsection{Heuristics Validation Attempt}
\label{sec:heuristic-validation}

Our heuristics in Section~\ref{sec:heuristics_all} are best-effort methods and may yield false positives and negatives, a known challenge of related works~\cite{androulaki2013evaluating, victor2020address, romiti2020cross}. To validate our heuristics, we observe the existence of a variety of publicly available side-channels that may indicate whether two blockchain addresses belong to the same entity. In this following, we expand on three of such side-channels, and then synthesize a candidate ground truth dataset to validate the results presented in Section~\ref{sec:heuristics_all}.

\subsubsection{Airdrop Side-Channel}
A blockchain airdrop is a form of donation, where a coin is given to a blockchain address without further explicit expectation. Victor et al.\ \cite{victor2020address} present the following privacy-related airdrop approach: if a user receives an airdrop on multiple addresses and aggregates those funds within a short timeframe after the airdrop to one central address, this address can be labeled as the user's primary address. As such the first side-channel we consider is the \emph{Airdrop} approach.

In our evaluation, we consider two particular instances of \DeFi airdrops: the \href{https://uniswap.org/blog/uni}{Uniswap airdrop} and  \href{https://airdrops.io/1inch/}{1inch airdrop}. To apply Victor's heuristic, we crawl transaction data on the Ethereum network in the first seven days after an airdrop took place. 

\noindent\textbf{Results.} From the airdrop data, we identify a total of $35{,}081$ linked address pairs ($\mathcal{S}_{\texttt{Airdrop}}$).


\subsubsection{\ENS Side-Channel}
In the following, we propose two novel approaches to link addresses using \ENS~\cite{xia2021ethereum} data. \ENS is a decentralized naming service on Ethereum, aiming to map human-readable names (e.g., ``alice.eth'') to blockchain addresses. Similar to DNS, \ENS supports dot-separated hierarchical domains, and a domain owner can create subdomains (e.g., ``foo.alice.eth''). To map a new $\texttt{name}$ to an address $\addr$, a user registers the $\texttt{name}$ with $\addr$ and sets its expiry time. Users can also transfer the ownership of a name to another address, or assign subdomains to addresses. 


\noindent\textbf{Linking Addresses through \ENS Usage.} To cluster \ENS addresses, we provide two approaches:

\noindent\emph{Name Ownership Transfers}: Given two addresses $\addr_1$ and $\addr_2$, if $\addr_1$ transfers the ownership of an \ENS $\texttt{name}$ to $\addr_2$, before $\texttt{name}$ expires, and $\addr_1$ only transfers its name once, then $\link(\addr_1, \addr_2) = 1$.

\noindent\emph{Subdomain Assignments}: For addresses $\addr_1$ and $\addr_2$, if $\addr_1$ has an \ENS $\texttt{name}$ and assigns a subdomain of $\texttt{name}$ to $\addr_2$, then $\link(\addr_1, \addr_2) =~1$.


\noindent\textbf{Results.} To apply the \emph{Name Ownership Transfers} approach, we crawl all ($372{,}756$) \texttt{Transfer} events of the ENS registry contract until November 1st, 2021. We extract the address pairs $(\addr_1, \addr_2)$, where $\addr_1$ transfers a name to $\addr_2$ and $\addr_1$ only transfers its name once. This approach can link $4{,}399$ address pairs. To apply the \emph{Subdomain Assignments} approach, we crawl all~($900$) \texttt{NewOwner} events emitted when a user directly calls the \ENS registry contract. We then extract the address pairs $(\addr_1, \addr_2)$, where $\addr_1$ assigns subdomains to $\addr_2$. We can identify $725$ linked address pairs. In total, from the \ENS data, we can link $5{,}124$ address pairs, denoted as $\mathcal{S}_{\texttt{ENS}}$.

\subsubsection{Debank Side-Channel} \href{https://debank.com/}{Debank} is an online blockchain explorer for tracking DeFi user portfolios. Users can log into Debank through a wallet (e.g., MetaMask) and \emph{follow} other addresses, similar to a social network. We hence assume that a user is unlikely to follow its own addresses and propose the following approach. Note that this is the first side-channel we consider which yields a \emph{negative signal} on whether two addresses are linked.

\noindent\emph{Debank Following Relationship}: Given two addresses $\addr_1$ and $\addr_2$, if $\addr_1$ follows $\addr_2$, or $\addr_1$ is followed by $\addr_2$ on Debank, then $\link(\addr_1, \addr_2) \not= 1$.

\noindent\textbf{Results.} For each \TC depositor and withdrawer address, we crawl their follower and following addresses on Debank before November 1st, 2021, i.e.\ those Debank addresses that follow or followed by TC addresses. Out of $54{,}504$ TC addresses, we find that $655+258=913$ ($1.8$\%) addresses have at least one follower or following address on Debank. Let $\mathcal{S}_{\texttt{Debank}}$ be the set of \TC depositor-withdrawer pairs $(\addr_d, \addr_w)$, where $\addr_d$ follows $\addr_w$, or $\addr_d$ is followed by $\addr_w$ on Debank. Our results show that $|\mathcal{S}_{\texttt{Debank}}| = 150$, i.e., $150$ depositor-withdrawer pairs have a follower or following relationship.

\subsubsection{Validation Attempt}
In the following, we attempt to validate the heuristics presented in Section~\ref{sec:heuristics_all}, using $\mathcal{S}_{\texttt{Airdrop}}$, $\mathcal{S}_{\texttt{ENS}}$ and $\mathcal{S}_{\texttt{Debank}}$ as the candidate ground truth data. Note that we can only validate the link of \TC address pairs, not the link among deposit and withdrawal transactions. We, therefore, omit Heuristic $1$ from the validation process, as H1 does not link addresses.

\noindent\textbf{$\mathcal{S}_{\texttt{Airdrop}}$ + $\mathcal{S}_{\texttt{ENS}}$.} Table~\ref{tab:validation-attempt} shows the results of our heuristic validation by applying the side-channels given by $\mathcal{S}_{\texttt{Airdrop}}$ and $\mathcal{S}_{\texttt{ENS}}$. Unfortunately, $\mathbf{H}_2$, $\mathbf{H}_4$, and $\mathbf{H}_5$ appear to perform rather poorly, when compared to $\mathbf{H}_3$. This result appears plausible, when considering that $\mathbf{H}_3$ focuses on asset-transfers, which also applies to the Airdrop and \ENS side-channel data. Luckily, heuristic $3$ is the most potent heuristic to reduce the anonymity set size.



\noindent\textbf{Airdrop and ENS Side-Channel Intersection.}
To increase our confidence in the side-channel data, we intersect the candidate ground truth data sources: if an address pair $(\addr_1, \addr_2)$ is linked \emph{both} in $\mathcal{S}_{\texttt{Airdrop}}$ and $ \mathcal{S}_{\texttt{ENS}}$, then $\addr_1$ and $\addr_2$ are \emph{more} likely to be controlled by the same user. Nevertheless, the overlap size between the airdrop and the \ENS data consists of only $13$ pairs. We hence refrain from applying the intersected side-channel dataset to validate $\mathcal{S}_{\texttt{TC}}$.

\noindent\textbf{$\mathcal{S}_{\texttt{Debank}}$.}
We find that, out of the {$150$} depositor-withdrawer pairs in $\mathcal{S}_{\texttt{Debank}}$, {$34$ ($23\%$)} pairs are linked through Heuristics $2$, $3$, and $5$. Therefore, if we regard the Debank follower relationship data as the ground truth, then those {$34$} addresses cannot be owned by the same user; thus, we consider them false positives.


\noindent\textbf{Validation Results Summary.}
In conclusion, by applying the airdrop and \ENS side-channels as candidate ground truth datasets, our heuristics can achieve an average F1 score of $0.55$ (cf.\ Table~\ref{tab:validation-attempt}), whereas Heuristic~$3$ provides the strongest signal. Our results suggest that validating the heuristics presented in Section~\ref{sec:heuristics_all} is a challenging, but feasible task. Our results can be further extended with additional side channels to synthesize a larger candidate ground truth dataset (e.g., by crawling Twitter data from testnet wallet validations, additional blockchain explorer labels, etc.).

\begin{figure}[t]
    \centering
    \includegraphics[width=\columnwidth]{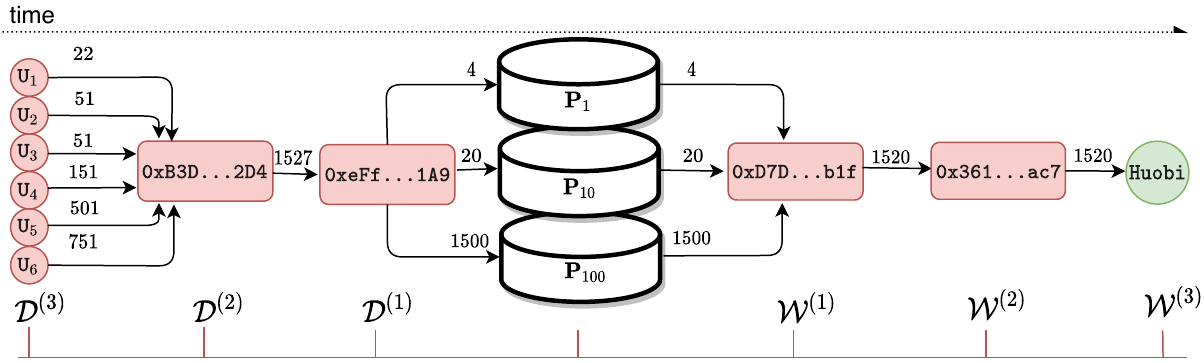}
    \caption{Applying linking heuristics to trace Upbit Hackers.}
    \Description{Applying linking heuristics to trace Upbit Hackers.}
    \label{fig:trace_upbit_hacker}
    \end{figure}

\fixed{
\section{Discussion and Implications}\label{sec:trace-address}



Our analyses show that although users may reveal their transaction history because they are not familiar with the workflow of mixers (cf.~Section~\ref{sec:measuring-as}), or they only use mixers for rewards rather than privacy (cf.~Section~\ref{sec:incentivisedpools}), most of the users can still stay anonymous. Our approach can be generalized to analyze any other \ZKP mixers which adopt the same design as \TC, e.g., \href{https://cyclone.xyz/bsc}{Cyclone} and \href{https://app.typhoon.network/}{\TP}. 

To improve the existing \ZKP mixer design, a helpful functionality could be to warn users proactively about potential risks. For example, TC could exploit our methodology and results to provide a service that would compute the probability that a provided address for a withdrawal could be linked with a depositor. In this way, users would know the risk of linking their addresses before withdrawing funds from the mixer. Moreover, besides the anonymity set size and the \OFAC sanctions, there might be other potential factors (e.g., \AM profits and \ETH or \BNB prices) which could also affect the usage of \ZKP mixers. We leave the detailed analysis for future work.
}

\subsection{Application: Tracing Malicious Addresses}\label{app:tracing-hackers}

We provide the example of Upbit Hackers to show how to apply our linking results in \TC to trace malicious addresses. On November 27th, 2019, hackers stole $342$K~\ETH from Upbit, a South-Korean centralized cryptocurrency exchange. As shown in Fig.~\ref{fig:trace_upbit_hacker}, \emph{(1)} A depositor \href{https://etherscan.io/address/0xeFf67710a1aE67885f660a965B0A8697CDB161A9}{\texttt{0xeFf}} receives $1{,}526.95$~\ETH from address \href{https://etherscan.io/address/0x5a88a3aD66234861621e983A498948Ffe6641857}{\texttt{0x5a8}}, which obtains the same amount of~\ETH from four labeled Upbit Hackers. \emph{(2)} \href{https://etherscan.io/address/0xeFf67710a1aE67885f660a965B0A8697CDB161A9}{\texttt{0xeFf}} then deposits $1{,}524$~\ETH into TC~1,~10, and~100~\ETH pools before block $11{,}972{,}040$. \emph{(3)} From our linking results, we find that \href{https://etherscan.io/address/0xD7D08d621c125e0131689839639C52E714038b1f}{\texttt{0xD7D}} withdraws the same amount from TC during block $11{,}971{,}270$ and $11{,}972{,}098$, and then transfers $1{,}520$~\ETH to address \href{https://etherscan.io/address/0x36164B276EA6F7B1C00d8E27d4E7dC8f28035ac7}{\texttt{0x361}}, which finally exchanges all \ETH to fiat currency (e.g., USD) on a \CEX, Houbi. Given the address's registration information on Huobi, it would be able to pinpoint the hacker's off-chain identity.

\section{Related Work}\label{sec:relatedwork}
    \noindent\textbf{Mixers on Bitcoin:} Mixers were originally applied in anonymous communications~\cite{chaum1981untraceable} and are also applied to enhance Bitcoin users' privacy~\cite{pakki2021everything,wu2021towards}. Mixcoin~\cite{bonneau2014mixcoin} and Blindcoin~\cite{valenta2015blindcoin} are centralized, trusted mixers that support \BTC.
    CoinJoin~\cite{maxwell2013coinjoin} allows a user to find other mixing partners to merge multiple transactions, thereby obfuscating the link between senders and recipients. Although the design of CoinJoin~\cite{maxwell2013coinjoin} is decentralized, its existing implementation, remains centralized but non-custodial. CoinShuffle~\cite{ruffing2014coinshuffle,ruffing2017coinshufflepp} and Xim~\cite{bissias2014Xim} achieve better anonymity in a decentralized mixer. Wu~\etal~\cite{wu2021towards} propose a generic abstraction model for Bitcoin mixers.

\noindent \textbf{Mixers on Smart-contract-enabled Blockchains:} \ZKP mixers are inspired by Zerocash~\cite{sasson2014zerocash} to obfuscate the link between the users' deposit and withdrawal using zero-knowledge proof. Several \ZKP mixers attempt to operate on Ethereum, such as Miximus~\cite{miximus}.  AMR~\cite{le2020amr} proposes how to reward users for participating in a mixer, and shortly after, \href{https://ourblender.github.io/}{Blender} implements a mixer with a reward scheme. \TC follows by adding anonymity mining as a deposit reward scheme for users~\cite{TornadoCashGovernanceProposal}. Besides \ZKP mixers, a notable mixer example that relies on linkable ring signatures and the stealth addresses from Monero~\cite{zero-to-monero} is Möbius~\cite{meiklejohn2018mobius}.

\noindent \textbf{Blockchain Privacy Analysis:} Many researchers have studied privacy on non-privacy-preserving blockchains (e.g., Bitcoin~\cite{androulaki2013evaluating,gervais2014privacy}, Ethereum~\cite{beres2021blockchain, victor2020address}), as well as on privacy-preserving blockchains (e.g., Monero~\cite{kumar2017traceability,moser2018empirical, yu2019new}, Zerocash~\cite{kappos2018empirical,biryukov2019privacy}). Because \ZKP mixers are inspired by Zerocash, our Heuristics $1$ and $4$ can also be applied to link shielded and deshielded transactions in Zerocash~\cite{kappos2018empirical}. However, the majority of the transactions (i.e., with $65.6\%$ of the withdrawn value) in~\cite{kappos2018empirical} involve miners or founders, while this paper investigates generic \ZKP mixers, and can be applied to trace malicious addresses. Moreover, recent studies~\cite{yousaf2019tracing} have shown that users' privacy can be leaked when using cross-chain exchanges.

\section{Conclusion}\label{sec:conclusion}
This paper empirically analyzes the usage of \ZKP mixers. We find that \numtotalmixerAttackers malicious addresses and \numtotalmixerBEVextractors \BEV extractors leverage mixers as their source of funds, while depositing a total attack revenue of \totalDepositofAttackersAndBEVextractors USD. We measure that the \OFAC sanctions have reduced more than {$83\%$} daily deposits in \TC. Moreover, our findings show that the advertised anonymity set sizes of popular mixers do not represent the true privacy offered to users. We propose a methodology that can reduce the anonymity set size on average by \totalTCASReduced (\totalTNASReduced) of \TC (on \ETHchain) and \TN (on \BSChain) respectively. Worryingly, while previous work suggests that incentivized mixers could improve the offered mixer privacy, we find evidence that speculators are likely to act in a privacy-ignorant manner, deteriorating the overall anonymity set size.
We hope that our work engenders further research into user-friendly and privacy-enhancing \ZKP mixer solutions.



\begin{acks}
We thank the anonymous reviewers for providing valuable comments and feedback which helped us to strengthen the paper. We are moreover grateful to Nimiq for partially funding this work.
This work was partially supported by the Algorand Centres of Excellence programme managed by Algorand Foundation. Any opinions, findings, and conclusions or recommendations expressed in this material are those of the author(s) and do not necessarily reflect the views of Nimiq and Algorand Foundation.
\end{acks}

\bibliographystyle{ACM-Reference-Format}
\bibliography{references}

\clearpage

\end{document}